\documentclass[a4paper,fleqn,usenatbib]{mnras}

\usepackage{amsmath}	% Advanced maths commands
\usepackage{amssymb}	% Extra maths symbols

\usepackage{mathptmx}
\usepackage{txfonts}

\usepackage[T1]{fontenc}
\usepackage{ae,aecompl}

\usepackage{float}

\usepackage{graphicx}	% Including figure files

\newcommand{\Hr}{\mathrm{H}}
\newcommand{\kms}{km~s$^{-1}$}
\newcommand{\Qd}{Q_\mathrm{D}^{\star}}
\newcommand{\dotM}{\dot{\mathrm{M}}}
\newcommand{\nel}{$n_{\mathrm{e}^{-}}$}

\newcommand{\etacorvi}{$\eta$~Corvi}
\title[ALMA observations of the $\eta$ Corvi debris disc]{ALMA
  observations of the $\eta$ Corvi debris disc: inward scattering of
  CO-rich exocomets by a chain of 3-30 M$_\oplus$ planets?}

\author[]{S. Marino$^{1}$\thanks{E-mail: s.marino@ast.cam.ac.uk}, M. C. Wyatt$^{1}$, O. Pani\'{c}$^{1,2}$\thanks{Royal Society Dorothy Hodgkin Fellow}, L. Matr\`{a}$^{1}$, G. M. Kennedy$^{1}$, A. Bonsor$^{1}$, 
  \newauthor{Q. Kral$^{1}$, W. R. F Dent$^{3}$, G. Duchene$^{4,5}$,  D. Wilner$^{6}$, C. M. Lisse$^{7}$, J.-F. Lestrade$^{8}$} \newauthor{and B. Matthews$^{9}$.} 
  \\
  $^{1}$Institute of Astronomy, University of Cambridge, Madingley Road, Cambridge CB3 0HA, UK\\
  $^{2}$School of Physics and Astronomy, University of Leeds, Woodhouse Lane, Leeds, LS2 9JT, UK\\
  $^{3}$Joint ALMA Observatory, Alonso de C\'ordova 3107, 763-0355 Vitacura, Santiago, Chile\\
$^{4}$Department of Astronomy, UC Berkeley, Berkeley, CA 94720, USA\\
$^{5}$Univ. Grenoble Alpes/CNRS, IPAG, F-38000 Grenoble, France\\
$^{6}$Harvard-Smithsonian Center for Astrophysics, 60 Garden Street, MS-78, Cambridge, MA 02138, USA\\
$^{7}$Space Exploration Sector, Johns Hopkins University Applied Physics Laboratory, 11100 Johns Hopkins Road, Laurel, MD, 20723, USA\\
$^{8}$Observatoire de Parris - LERMA, CNRS, 61 Av. de l\'Observatoire, 75014 Paris, France\\
$^{9}$National Research Council of Canada Herzberg Astronomy and Astrophysics Programs, 5071 West Saanich Road, Victoria, BC, V9E 2E7,\\ Canada
}

\date{Accepted 2016 November 3. Received 2016 November 2; in original form 2016 October 3}

\pubyear{2016}

\begin{document}
\label{firstpage}
\pagerange{\pageref{firstpage}--\pageref{lastpage}}
\maketitle

% Abstract of the paper
\begin{abstract}

While most of the known debris discs present cold dust at tens of AU,
a few young systems exhibit hot dust analogous to the Zodiacal
dust. \etacorvi \ is particularly interesting as it is old and it has
both, with its hot dust significantly exceeding the maximum luminosity
of an in-situ collisional cascade. Previous work suggested that this
system could be undergoing an event similar to the Late Heavy
Bombardment (LHB) soon after or during a dynamical instability. Here
we present ALMA observations of \etacorvi \ with a resolution of
$1\farcs2$ ($\sim$22au) to study its outer belt. The continuum
emission is consistent with an axisymmetric belt, with a mean radius
of 152au and radial FWHM of 46au, which is too narrow compared to
models of inward scattering of an LHB-like scenario. Instead, the hot
dust could be explained as material passed inwards in a rather stable
planetary configuration. We also report a $4\sigma$ detection of CO at
$\sim20$au. CO could be released in situ from icy planetesimals being
passed in when crossing the H$_2$O or CO$_{2}$ ice lines. Finally, we
place constraints on hidden planets in the disc. If a planet is
sculpting the disc's inner edge, this should be orbiting at 75-100au,
with a mass of 3-30 M$_\oplus$ and an eccentricity $<0.08$. Such a
planet would be able to clear its chaotic zone on a timescale shorter
than the age of the system and scatter material inwards from the outer
belt to the inner regions, thus feeding the hot dust.

\end{abstract}

% Select between one and six entries from the list of approved keywords.
% Don't make up new ones.

%%%%%%%%%%%%%%%%%%%%%%%%%%%%%%%%%%%%%%%%%%%%%%%%%%

%%%%%%%%%%%%%%%%% BODY OF PAPER %%%%%%%%%%%%%%%%%%

\section{Introduction}
\label{sec:intro}
As a byproduct of planet formation, discs of km-sized bodies or
planetesimals can form, analogous to the Asteroid and Kuiper belts in
the Solar System \citep{Lissauer1993}. Destructive collisions between
these bodies can give rise to debris discs, grinding down the largest
bodies producing a wide size distribution of solids, from $\mu$m-sized
dust grains up to the size of asteroids or comets, in a so-called
collisional cascade \citep{Wyatt2007collisionalcascade}. This has been
proven observationally using multiwavelength observations \citep[see
][]{Backman1993}. Moreover, the last decade of debris disc surveys has
shown that Kuiper belt analogues are quite common, with detection
rates of at least 20\% around AFGK stars \citep{Su2006,
  Hillenbrand2008, Carpenter2009, Eiroa2013, Thureau2014,
  Matthews2014pp6}. These surveys have also shown that the levels of
infrared excess decay with stellar age, as expected from collisional
models.

By studying the spectral energy distribution (SED) of systems with
detected dust, it is possible to estimate the temperature and radial
location of the dust assuming standard optical properties \citep[e.g.,
][]{Kennedy2014}. While in most of the known debris discs the dust
grains are located in narrow belts at tens of AU, analogous to the
Kuiper belt, a small fraction of discs have been identified to host a
hot dust component within a few AU, analogous to the Asteroid belt or
Zodiacal dust \citep[e.g.,][]{Absil2013}. The majority of these discs
with hot dust are younger than $\sim100$ Myr old \citep{Kennedy2013},
e.g HD~172555 \citep{Lisse2009}.

Since collisional timescales are a steep function of the distance to
the star \citep{Lohne2008}, debris discs at a few AU are expected to
evolve much faster than Kuiper belt analogues. This implies that there
is a maximum possible disc mass at a given radius and age, and thus a
maximum dust luminosity \citep{Wyatt2007collisionalcascade,
  Lohne2008}. However, an important subset of discs with hot dust
around old ($\gg$100~Myr) stars exceed this limit by 3 orders of
magnitude or more, precluding its explanation as a result of
destructive collisions between planetesimals in situ
\citep{Wyatt2007hotdust}. Therefore, in most of these systems the
planetesimals feeding the hot dust component must be located farther
from the star, in a cold belt of planetesimals where collisions are
less frequent and the disc mass decays more slowly.

%% Moreover, the hot dust component can only be
%% explained by a population of eccentric planetesimals if their
%% eccentricities, $e$, are larger than 0.99
%% \citep{Wyatt2010}.

Among the sample of old stars with hot dust, the $1-2$ Gyr old F2V
star \etacorvi \ \citep[HD~109085, HIP~61174,][hereafter assumed to be
  1.4 Gyr old, consistent with isochrone fitting in the HR
  diagram]{Ibukiyama2002, Mallik2003, Vican2012}, located 18.3 pc away
\citep{vanLeeuwen2007}, is particularly interesting because it
presents both a hot and a cold dust component in its SED, where the
hot dust exceeds the brightness limit mentioned above. Its infrared
excess was first detected with \textit{IRAS} \citep{Stencel1991},
after which several observations from the optical to the millimetre
have targeted this system. From mid-infrared observations,
\cite{Smith2009etacorvi} determined that the hot dust must be located
within 3 AU from the star. More recent observations with the Large
Binocular Telescope Interferometer (LBTI) and the Keck Interferometer
have confirmed this and determined that the hot component is located
at projected separations of $\sim1.4$~AU and 0.2-0.8~AU from the star,
respectively \citep{Defrere2015, Kennedy2015lbti,
  Lebreton2016}. Moreover, we know that the the hot dust luminosity
has stayed almost constant over the last three decades
\citep{Duchene2014}. On the other hand, \cite{Wyatt2005} resolved the
cold component of the system with SCUBA/JCMT at millimetre
wavelengths, finding that it has a mean radius of $\sim150$ AU. At 70
$\mu$m \textit{Herschel} imaged both hot and cold components,
resolving the outer belt and confirming that the hot dust has large
amounts of small dust below the blow-out size \citep[previously known
  from \textit{IRS},][]{Chen2006} providing further evidence that the
hot dust cannot be explained by a steady-state collisional cascade and
suggesting a more violent origin \citep{Duchene2014}.

Several scenarios have been proposed to explain the high levels of
mid-infrared emission in \etacorvi. \cite{Wyatt2010} suggested that
the system could host a population of highly eccentric planetesimals,
colliding at pericentre within a few AU producing the hot dust and
observed excess. These planetesimals would survive for longer
timescales due to their large apocentre in the cold belt. While this
scenario is appealing in connecting the the inner and outer dust
components in the system, it was ruled out by \textit{Herschel}
resolved observations \citep{Duchene2014} as no emission is detected
between the hot dust location and the outer belt. The hot dust could
be also a product of a giant impact on a planet close in or recent
giant collision in an in situ asteroid belt rather than an ongoing
collisional cascade. This would be consistent with the spectroscopic
features of impact produced silica and high-temperature carbonaceous
phases revealed by \textit{Spitzer} \citep{Chen2006, Lisse2012}. While
these events are expected to be rare and the produced small dust short
lived compared to the age of the system \citep{Kral2015}, \etacorvi
\ is one of just two FGK stars with hot dust at this level out of the
DEBRIS sample of $\sim300$ stars, i.e. we cannot exclude an unlikely
scenario as we could be witnessing the system at a special
time. However, the hot dust also displays spectral features of
primitive cometary material consistent with icy solids formed further
out in the system \citep{Lisse2012}, and thus, transported by an
unknown process to within a few AU. There are two known mechanisms
that can transport material inwards, Poynting-Robertson (P-R) drag and
interactions with a single or multiple planets. The first can deliver
small grains from the outer disc to the inner regions, but this
mechanism is not efficient enough to explain the hot dust in
\etacorvi, given the low optical depth or fractional luminosity of the
outer belt \citep{Kennedy2015prdrag}. Thus, it seems necessary to
invoke the presence of planets in the system.

Therefore, three scenarios remain that could explain the hot dust
level and its composition: 1) the system is going through an
instability, analogous to the Late Heavy Bombardment (LHB) in the
Solar System, scattering comets from the outer belt to the innermost
regions \citep[e.g.,][]{Bonsor2013}; 2) planets in a rather stable
configuration in the system are scattering dust from the outer disc
feeding the hot dust or bigger icy solids that then collide closer in
producing the hot dust \citep[e.g.,][]{Bonsor2012nbody}; 3)
planetesimals are scattered by planets in the system, colliding with a
planet within a few AU releasing large amounts of debris. Thus, we are
interested in studying the structure of the outer belt to look for
features that can hint at one of the three possible planet-driven
scenarios described above to explain the hot dust.

%% w being delivered to the inner regions colliding
%% with the Earth and Moon.

%% This was
%% suggested by \cite{Lisse2012}

%% spectroscopic studies with \textit{Spitzer} revealed that the hot dust
%% has features of primitive cometary material, as well as features of
%% impact produced silica and high-temperature carbonaceous phases
%% \citep{Chen2006, Lisse2012}.

In this paper we present the first continuum observations of \etacorvi
\ with the Atacama Large Millimeter/submillimeter Array (ALMA) to
study in detail the continuum dust emission of its cold outer belt at
0.88 mm. At this wavelength, the continuum is dominated by mm-sized
dust grains ($\sim0.1-10.0$ mm), for which radiation forces are
negligible. Thus, they trace the distribution of the parent
planetesimals which contain the bulk of the disc mass and may provide
clues to the mechanism feeding the hot dust closer in. In
Sec. \ref{sec:obs} we describe the ALMA observations, continuum
imaging and search for any CO emission in the disc. Then, in
Sec. \ref{sec:model} we model the continuum data with parametric disc
models to study the distribution of planetesimals in the disc. We also
derive CO gas masses and upper limits considering non-LTE. In
Sec. \ref{sec:dis} we discuss the implications of the ALMA continuum
data, how it fits with the hypothesis of material being transported
from the outer to the inner regions of the disc; we also discuss the
origin of the tentative CO gas detection, and we place some
constraints on a planet between the hot dust and outer belt based on
the continuum observations. Finally in Sec. \ref{sec:conclusions} we
summarise the main results and conclusions of our work.

\section{Observations}
\label{sec:obs}

\newlength\qa
\setlength\qa{\dimexpr .15\textwidth -2\tabcolsep}
\newlength\qb
\setlength\qb{\dimexpr .08\textwidth -2\tabcolsep}

% Example table
\begin{table*}
  \centering
  %\begin{adjustbox}{max width=\textwidth}
  \caption{Summary of science observations.}
%    Remember to define the quantities, symbols and units used.}
    \label{sciobs}
    \noindent\begin{tabular}{p{\qa}p{\qa}p{\qb}p{\qb}p{\qb}p{\qb}p{\qa}p{\qa}p{\qa}} % four columns, alignment for each
        \hline
Date of observations & elevation [deg] & n$_\mathrm{ant}$ & t$_\mathrm{sci}$ [min] & $\%$ flagged & image rms [$\mu$Jy] & Flux calibrator & Bandpass calibrator & Phase Calibrator \\
        \hline
        2013 Dec 15 & 51-61 & 27 & 40 & 16 & 90 & Pallas   & J1058-0133 & J1215-1731 \\
        2013 Dec 15 & 70-79 & 27 & 40 & 40 & 90 & Ceres    & J1256-0547 & J1215-1731 \\
        2014 Dec 25 & 72-83 & 38 & 48 & 20 & 60 & 3c279    & J1256-0547 & J1245-1616 \\
        2014 Dec 26 & 77-83 & 40 & 48 & 18 & 40 & Ganymede & J1256-0547 & J1245-1616 \\
        2014 Dec 29 & 66-81 & 37 & 48 & 15 & 50 & Titan    & J1256-0547 & J1245-1616 \\
        2015 Jan 01 & 62-77 & 37 & 48 & 20 & 50 & Titan    & J1256-0547 & J1245-1616 \\
                \hline
    \end{tabular}
  %\end{adjustbox}
\end{table*}

ALMA band 7 observations (340 GHz) of \etacorvi \ were carried out
from 2013 Dec 15 to 2015 Jan 1 under the project 2012.1.00385.S (PI:
M. C. Wyatt), resulting in 6 successful executions. Each set of
observations is a 3-point mosaic along the disc major axis, separated
by 8$\farcs$8 and centered on \etacorvi. The primary beam FWHM of a
single pointing is 17$\farcs$9, which allowed to reach a combined
primary beam efficiency above 50\% in a elongated region of
36$\arcsec$x18$\arcsec$ oriented as the disc in the sky. Array
configurations were very compact, with baselines ranging from 15~m to
445~m, with the 5th and 95th percentiles equal to 26~m and 240~m. This
allows to recover angular scales from $1\arcsec$ up to $7\arcsec$.

The band 7 spectral setup consisted of four spectral windows. Three
were set to observe the continuum emission centered at 335.744,
337.644 and 347.455 GHz, each one with 128 channels of 14~km/s width
(effective spectral resolution of 28~\kms), obtaining a 2 GHz total
bandwidth per spectral window. The fourth included the $^{12}$CO 3--2
line at 345.798~GHz and was set with 3840 finer channels, each one
with a width of 0.42~\kms (effective spectral resolution of
0.82~\kms), obtaining a total bandwidth of 1.9 GHz. The four together
provided an effective bandwidth of 7.9~GHz to study the continuum
emission.

Table ~\ref{sciobs} provides a summary of the six sets of science
observations of \etacorvi \ spanning approximately one year as the
ALMA array expanded from n$_\mathrm{ant}\approx$30 to $\approx$40
antennas. Consequently the most sensitive observations were the ones
obtained most recently. We can also see that the science target has
always been observed at a relatively high elevation. The total time
on-source t$_\mathrm{sci}$ shown corresponds to the sum of the three
pointings in the mosaic, so the actual time spent on-source with the
pointing on the location of \etacorvi \ itself is one third of
t$_\mathrm{sci}$. Pallas, Ceres, 3c279, Ganymede and Titan were used
as flux calibrators and observed for 2.6 min; J1058-0133 and
J1256-0547 as Bandpass calibrators and observed for 5.3 min; and
J1215-1731 and J1245-1616 as phase calibrators and observed for 10.9
and 14.1 min in total on each science observation, respectively.

%% Table~\ref{cals} summarises the calibrators used
%% for the 6 aforementioned science observations, along with their
%% elevations and time on source.

% Example table
%% \begin{table*}
%%     \centering
%%     \caption{Summary of calibrations.}
%% %    Remember to define the quantities, symbols and units used.}
%%     \label{cals}
%%     \begin{tabular}{llllllllll} % four columns, alignment for each
%%         \hline
%% Date of & Flux & elevation & t$_\mathrm{obs}$ & Bandpass & elevation & t$_\mathrm{obs}$ & Phase & elevation & t$_\mathrm{obs}$ \\
%% observation & calibrator & [deg] & [min] & calibrator & [deg] & [min] & calibrator & [deg] & [min] \\
%%                 \hline
%% 2013 Dec 15 & Pallas & 85 & 2.6 & J1058-0133 & 58 & 5.3 & J1215-1731 & 54-62 & 10.9 \\
%% 2013 Dec 15 & Ceres & 50 & 2.6 & J1256-0547 & 58 & 5.3 & J1215-1731 & 70-84 & 10.9 \\
%% 2014 Dec 25 & 3c279 & 72 & 2.6 & J1256-0547 & 71 & 5.3 & J1245-1616 & 78-84 & 14.1 \\
%% 2014 Dec 26 & Ganymede &42 & 2.6 & J1256-0547 & 63 & 5.3 & J1245-1616 & 74-83 & 14.1 \\
%% 2014 Dec 29 & Titan & 42 & 2.6 & J1256-0547 & 63 & 5.3 & J1245-1616 & 74-83 & 14.1 \\
%% 2015 Jan 01 & Titan & 51 & 2.6 & J1256-0547 & 73 & 5.3 & J1245-1616 & 65-80 & 14.1 \\
%%                 \hline
%%     \end{tabular}
%% \end{table*}

\etacorvi \ has a significant proper motion of $-425.2\pm0.2$ and
$-57.2\pm0.1$~mas/yr in RA and Dec direction respectively
\citep{vanLeeuwen2007}, and consequently the pointing of the mosaic
had to be adjusted for each observation. Unfortunately, due to an
identified and subsequently corrected error in carrying out mosaic
observations\footnote{Described in
  https://almascience.eso.org/news/notification-of-problem-affecting-certain-alma-data-that-used-mosaic-and-offset-pointing-observing-modes}
the pointings were not always consistently centered on the star and
neither were the offsets always equal. These errors were corrected
using the CASA task fixvis \citep{McMullin2007} and subsequently
aligning and combining the data prior to the image
synthesis. Deviations of up to +0.36~arcsec in RA and +0.48~arcsec in
Dec were made. This is sufficiently small compared to the FWHM of the
primary beam to allow to successfully combine the observations,
achieving the desired rms.

\subsection{Continuum emission}
\label{sec:continuum}

\begin{figure}
  \includegraphics[trim=0.5cm 1.5cm 0.5cm 0.5cm, clip=true,
    width=1.0\columnwidth]{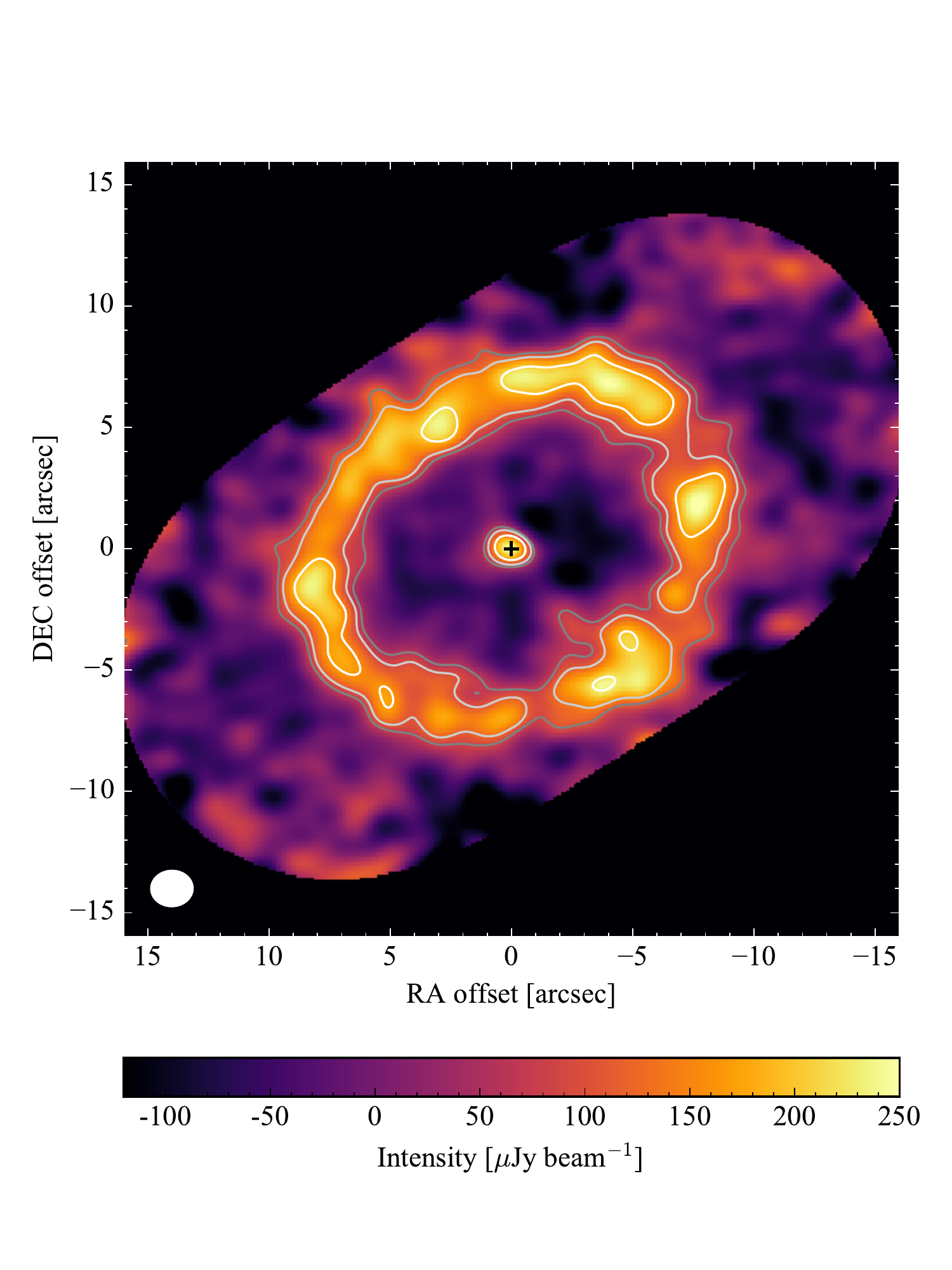}
  \caption{ALMA dust continuum CLEAN image at 0.88 mm (Band 7), with
    natural weights, primary beam corrected, and tapered by a Gaussian
    of FWHM of $1\farcs2$. The beam size is $1\farcs5\times1\farcs3$
    and is represented by a white ellipse in the bottom left
    corner. At the center of the image, the noise level is 30
    $\mu$Jy~beam$^{-1}$. The black masked region indicates a primary
    beam response below 40\%. The contours represent emission above 2,
    3 and 5 times the local noise level. The x- and y-axes indicate
    the offset from the stellar position in R.A. and decl. in arcsec,
    i.e. north is up and east is left. The stellar position is marked
    with a black ``+''. }
    \label{fig:cleannatural}
\end{figure}

The image reconstruction of the dust continuum is performed with the
\textit{CLEAN} algorithm and task in CASA 4.4 \citep{McMullin2007},
using natural weights, and combining the four spectral windows to
recover the highest S/N. As the dirty image of the data has strong
negative artefacts between the stellar emission and the outer belt, we
first clean the stellar emission and then the outer belt. Using
natural weights, we obtain an elongated beam of size
$1\farcs18\times0\farcs65$ (equivalent to $22\times12$ AU), with its
major axis oriented with a position angle (PA) of 89$^{\circ}$. The
image noise level before correcting by the primary beam is 20
$\mu$Jy~beam$^{-1}$ (which does not take into account the uncertainty
on the absolute flux calibration, $\sim$10\%). This is estimated
measuring the image standard deviation in a elliptic annulus far from
the disc emission. Finally we corrected by the primary beam obtaining
an image noise that increases with distance to the star as the primary
beam efficiency decreases. The image noise around the outer belt
varies from 22~$\mu$Jy~beam$^{-1}$ near the disc major axis up to
30~$\mu$Jy~beam$^{-1}$ close to the minor axis. In Figure
\ref{fig:cleannatural} we present the CLEAN image tapering the
visibilities with the Fourier transform of a Gaussian of FWHM of
$1\farcs2$, to obtain a less elongated beam of
$1\farcs5\times1\farcs3$ with a PA of $88^{\circ}$, and higher S/N per
beam. At the center of the tapered image, the noise level is
30~$\mu$Jy~beam$^{-1}$, varying from 33 to 45~$\mu$Jy~beam$^{-1}$
around the outer belt. Substructure is present along the belt in the
Clean image, with two minima to the south west of the star. However,
the emission is still consistent with an axisymmetric belt given the
S/N (see Sec. \ref{sec:model}). Compared with the PACS image at 70
$\mu$m obtained by \textit{Herschel} (previous best image of the outer
belt), we obtain a spatial resolution four times higher or a 50\%
higher S/N if we degrade the ALMA image to the PACS resolution.

%% In Figure \ref{fig:cleannatural} we present the deconvolved CLEAN
%% image using natural weights obtained using the software CASA 4.4
%% \citep{McMullin2007} and combining the four spectral windows to
%% recover the highest S/N. As the dirty image of the data has strong
%% negative artefacts between the stellar emission and the outer belt, we
%% first clean the stellar emission and then the outer belt. Using
%% natural weights, we obtain an elongated beam of size
%% $1\farcs18\times0\farcs65$ (equivalent to $22\times12$ AU), with its
%% major axis oriented with a position angle of 89$^{\circ}$. The image
%% noise level before correcting by the primary beam is 20
%% $\mu$Jy~beam$^{-1}$ (which does not take into account the uncertainty
%% on the absolute flux calibration, $\sim$10\%), obtained measuring the
%% image standard deviation in a elliptic annulus far from the disc
%% emission. Finally we corrected by the primary beam obtaining an image
%% noise that increases with distance to the star as the primary beam
%% efficiency decreases. The image noise around the outer belt varies
%% from 22~$\mu$Jy~beam$^{-1}$ near the disc major axis up to
%% 30~$\mu$Jy~beam$^{-1}$ near the minor axis.

Using the non-tapered image, we obtain an intensity peak of $258\pm20$
$\mu$Jy~beam$^{-1}$ at the stellar location, which is consistent with
the expected photospheric emission of $250\pm7$ $\mu$Jy, extrapolated
from the available photometry at 2.2, 3.6, 3.8, and 4.8 $\mu$m
assuming a spectral index of 2 \citep{Sylvester1996}. However, the
stellar emission at these wavelengths could deviate significantly from
Rayleigh-Jeans as observed for the Sun
\citep[e.g.,][]{Loukitcheva2004, Fontenla2007}. Subtracting the
extrapolated stellar flux we can place an upper limit on any extra
emission, e.g. hot dust. Considering the different uncertainties we
find a $3\sigma$ upper limit of 100 $\mu$Jy on the hot dust flux at
0.88 mm, higher than the 44$\mu$Jy extrapolated from the 70~$\mu$m
flux excess detected with \textit{Herschel} \citep{Duchene2014} using
a spectral index of 2.0 (optimistic prediction), and far above the
1.5$\mu$Jy predicted by the hot dust model presented in the same study
to fit the SED. Therefore, it is not surprising that no hot dust
emission is detected.

We measure a total disc flux of $9.2\pm0.5$ mJy, integrating the
continuum emission inside an elliptic mask with the same orientation
and ellipticity as the disc, i.e. inclination ($i$) of $35^{\circ}$
and a position angle (PA) of $117^{\circ}$ (the estimation of $i$ and
PA is presented in Sec. \ref{sec:model}, obtaining values consistent
with previous Herschel observations, \citealt{Lebreton2016}), and with
semi-major axis of $11\arcsec$. However, we notice that there is
significant negative emission inside the outer belt. These negatives
may occur when deconvolving visibility data using the CLEAN image
reconstruction algorithm, which does not impose positivity. If we
integrate the flux inside an elliptic annulus of minimum and maximum
semi-major axes $6\arcsec$ and $11\arcsec$, we obtain a higher
integrated flux of $10.1\pm0.4$ mJy, where the uncertainties stated
above only consider the image noise.

At a similar wavelength of 0.85 mm, SCUBA-2/JCMT observed \etacorvi
\ measuring a total flux of $15.5\pm1.4$ mJy. Extrapolating this to
0.88 mm with a spectral index of 3.0, i.e. $14.0\pm1.3$ mJy, we find
that the total emission measured in the ALMA Clean image is lower by
$2-3\sigma$ considering absolute flux calibration uncertainties. The
difference could be due to extended emission missed by an insufficient
number of short baselines or due to the image reconstruction
method. In Sec. \ref{sec:model} we model the disc emission by fitting
different disc models to the visibilities, and thus we obtain disc
fluxes that do not rely in the image deconvolution method. Fitting a
simple axisymmetric belt we find a total flux of $13.3\pm1.6$,
consistent with the total flux measured by SCUBA-2/JCMT at a similar
wavelength.

As mentioned above we find some non-negligible artefacts in the
reconstructed images of the data. These appear both in the dirty and
CLEAN images. These could be caused by: (1) extended emission not
captured by the uv-coverage due to insufficient short baselines; (2)
the gridding of the visibilities when computing the fast Fourier
transform, which can produce a dirty image that does not approximate
well to the true sky brightness for our uv-coverage; (3) imprecisions
in the uv coordinates due to pointing errors when carrying out the
mosaic observations as mentioned before. Because the negatives appear
both at the real and simulated observations (see
Sec. \ref{sec:model}), as well as using a different gridding or a
uv-sampling when simulating observations, it is unlikely that the
negatives are caused by (2) and (3). This shows the importance of
having a complete uv-coverage that includes baselines that can capture
the angular size of the observed source, e.g. including the 7m array
(ACA) in ALMA observations or combining with SMA
observations. Moreover, in such cases of incomplete uv-sampling,
e.g. insufficient number of short baselines, models should be compared
with interferometric observations in the visibility space as
reconstructed images can suffer from non-negligible artefacts.

%% \begin{figure}
%%   \includegraphics[trim=1.5cm 0.5cm 1.0cm 0.5cm, clip=true,
%%     width=1.0\columnwidth]{Figure_continuum_rainbow_pbcor.png}
%%   \caption{ALMA dust continuum CLEAN image at 0.88 mm (Band 7), with
%%     natural weights and primary beam corrected. The beam size is
%%     $1\farcs18\times0\farcs65$ and is represented by a white ellipse
%%     in the bottom left corner. The image has a noise level of 20
%%     $\mu$Jy~beam$^{-1}$. The black masked region indicates a primary
%%     beam response below 20\%. The x- and y-axes indicate the offset
%%     from the stellar position in R.A. and decl. in arcsec, i.e. north
%%     is up and east is left. The stellar position is marked with a
%%     black ``+''. }
%%     \label{fig:cleannatural}
%% \end{figure}

\begin{figure}
  \includegraphics[trim=0.0cm 2.0cm 0.0cm 0.0cm, clip=true,
    width=1.0\columnwidth]{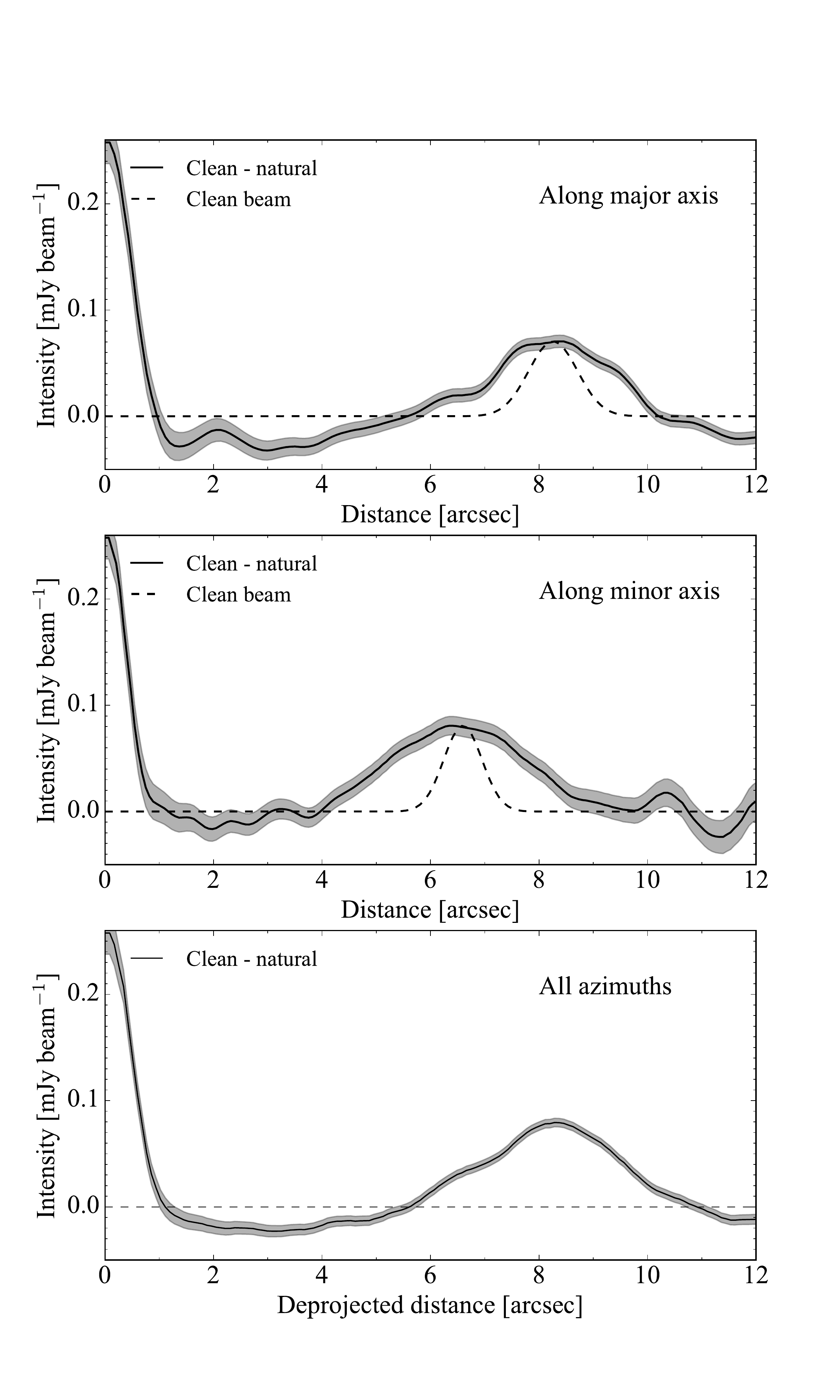}
  \caption{Intensity radial profiles of the dust continuum vs distance
    to the star along the major (top panel) and minor axis of the disc
    (middle panel) obtained averaging over 40$^{\circ}$ wide
    wedges. In the bottom panel we present the mean intensity at all
    azimuths vs the deprojected distance to the star. The grey areas
    represent the 68\% confidence region.}
    \label{fig:Ir}
\end{figure}

To study the radial profile of the surface brightness we compute the
average intensity over $40^{\circ}$ wide wedges along the major and
minor axes, and the average intensity profile at all azimuths around
ellipses oriented as above. Each point in the intensity profiles
corresponds to the mean intensity over 100 points equidistant in
position angle around an ellipse with a fixed semi-major axis or
distance. The intensity at each point around the ellipses is
calculated based on the value of the nearest pixel of the non-tapered
CLEAN image. This is analogous to using elliptic annuli with radial
widths of $0\farcs1$ (pixel size of the deconvolved image). Finally,
the uncertainty of the radial profile is calculated as the squared
root of the mean squared rms along each ellipse, divided by the square
root of the number of independent points around the ellipse,
approximated as the perimeter of the ellipse divided by the beam major
axis. Figure \ref{fig:Ir} shows the intensity profiles. We stress
that points in the radial profile separated by less than
$\sim1\arcsec$ (beam size) are not independent. The peak intensity
along the major and minor axes are $70\pm8$ $\mu$Jy~beam$^{-1}$ and
$81\pm12$ $\mu$Jy~beam$^{-1}$, respectively, both consistent within
the uncertainties, while the average profile at all azimuths has a
peak intensity of $79\pm4$ $\mu$Jy~beam$^{-1}$. Along the major axis
of the disc, the belt peaks at $8\farcs30\pm 0\farcs05$ ($151\pm1$
AU), where the error is estimated as $\sim$Beam$_\mathrm{fwhm}/(S/N)$,
with S/N$\sim26$ at the peak -- this uncertainty could be
underestimated as the relation above is strictly valid only for point
sources. The deprojected radial profile in the lower panel of Figure
\ref{fig:Ir} shows that the belt is clearly spatially resolved,
spanning over $\sim4\arcsec$, or $\sim70$~AU in radial extent.

\begin{figure}
  \includegraphics[trim=0.0cm 0.0cm 0.0cm 0.0cm, clip=true,
    width=1.0\columnwidth]{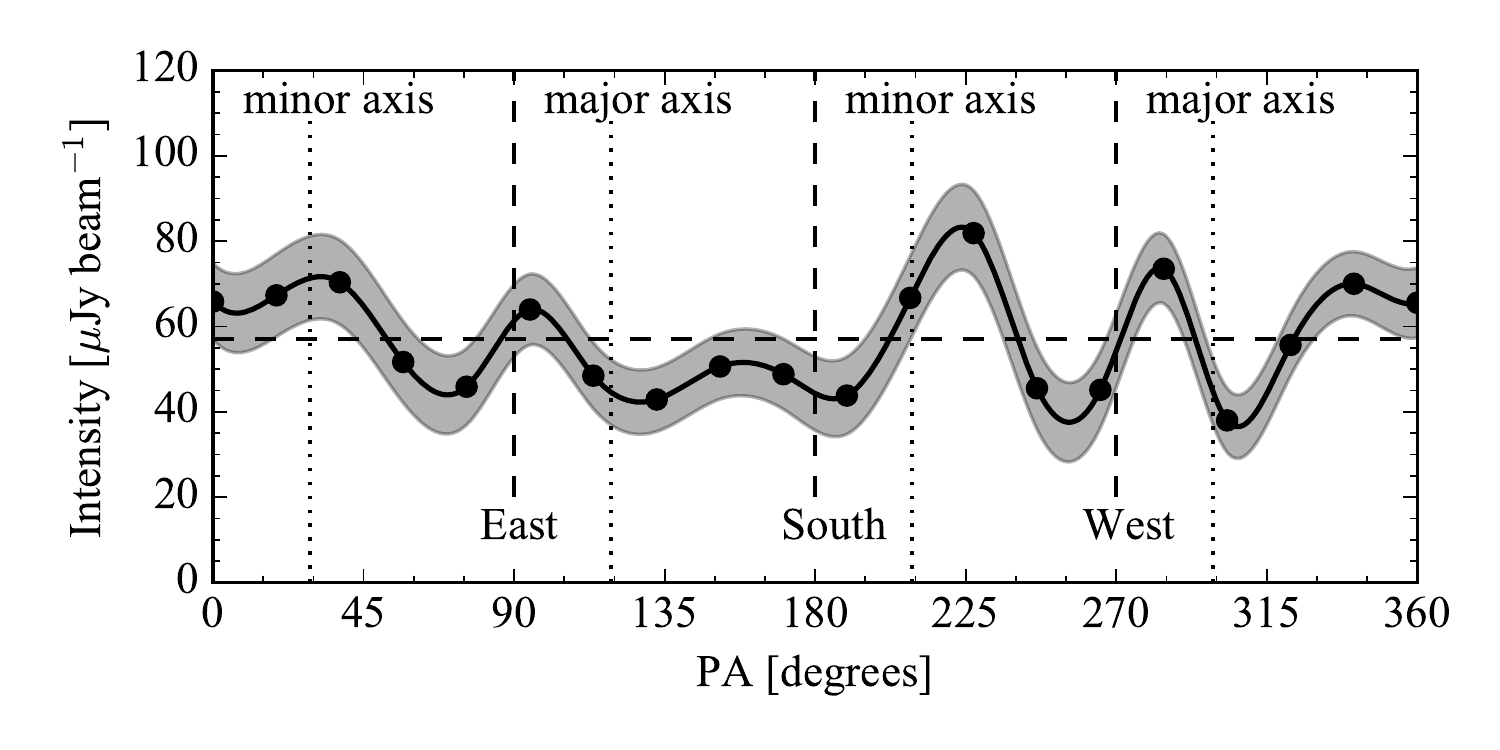}
  \caption{Intensity azimuthal profiles of the dust continuum emission
    vs position angle (PA) obtained averaging over elliptic annuli of
    semi-major axes between $6\farcs8$ and $9\farcs8$ (equivalent to
    125 and 180 AU) and over 18$^{\circ}$ wide wedges (centered at the
    black dots). The continuous line is obtained with a cubic spline
    interpolation between the black dots. The horizontal dashed line
    represents the mean intensity around the outer belt. The grey area
    represents the 68\% confidence region.}
    \label{fig:Iphi}
\end{figure}

Similarly, we study the surface brightness variations along position
angle averaging the disc emission radially between 120 and 180 AU, and
over arcs of 18$^{\circ}$. This is shown in Figure
\ref{fig:Iphi}. Unlike \cite{Duchene2014} where they found evidence of
asymmetric disc emission at 70 $\mu$m with the north-west side having
a peak intensity higher by a factor of 1.4 compared to the south-east
side (see their Figure 5), we find no evidence of such asymmetries
around the azimuthal profile of the belt. In fact, following the same
procedure as they did, we compare the peak intensity of the averaged
radial profile of the north-west with the south-east ansae, finding
that they are consistent within uncertainties (ratio of $0.8\pm0.1$
with the south-east being brighter) and the first cannot be brighter
by more than a factor of 1.2 ($3\sigma$ limit). However, if we study
the contribution to the total flux from the north-west and south-east
halves of the disc (divided by the minor axis of the disc), we find
that they have fluxes of $5.6\pm0.3$ and $4.5\pm0.3$ mJy,
respectively. This is marginal evidence for emission excess on the
north-west side of the disc as the Herschel showed, but still
consistent with being axisymmetric.

%% . unlike \cite{Duchene2014} where they found
%% evidence of asymmetric disc emission at 70 $\mu$m with the north-west
%% side having a peak intensity higher by a factor of 1.4 compared to the
%% south-east side (see their Figure 5), we find no evidence of
%% asymmetries around the belt. In fact, following the same procedure as
%% they did, we compare the peak intensity of the averaged radial profile
%% of the north-west with the south-east ansae, finding that they are
%% consistent within uncertainties (ratio of $0.8\pm0.1$ with the
%% south-east being brighter) and the first cannot be brighter by more
%% than a factor of 1.2 ($3\sigma$ limit).

%% in the but
%% we find no evidence of asymmetries along the ring.

\subsection{CO (3-2)}
\label{sec:co}

The increasing number of gas detections in young debris discs has
opened a debate over whether primordial gas can remain in the disc for
more than 10 Myr, or whether it can be of secondary origin, e.g. being
released by volatile-rich icy solids in the disc. After subtracting
the continuum emission from the visibilities using the CASA task
\textit{uvcontsub}, we search for any CO $J=3-2$ gas emission that
could be present in the system. We find no significant CO gas emission
along the continuum subtracted dirty channel maps (CLEAN is not
necessary as there is no significant emission in the dirty channel
maps). As shown by \cite{Matra2015} and \cite{Marino2016}, the
detection limits can be improved by azimuthally averaging the dirty
channel maps and integrating in frequency. Similar to the method
presented in the previous section, we calculate the mean CO intensity
at different radii azimuthally averaging the continuum subtracted
dirty channel maps around ellipses oriented as the outer belt (see
Sec. \ref{sec:modelring}). We also integrate the emission in frequency
or radial velocity (RV) between the minimum and maximum RV expected
from Keplerian rotation at each radii, and assuming the same
inclination as the outer belt (37$^{\circ}$). We restrict the maximum
RV to 6 km~s$^{-1}$, equivalent to the maximum RV at 10 AU from the
star as we cannot spatially resolve emission within 10 AU. The stellar
radial velocity, $v_\star$, is not very well constrained in this
system with measurements ranging between -2.8 to 1.8 \kms in the
heliocentric reference frame \citep{Gontcharov2006,
  Casagrande2011}. Therefore, we vary $v_\star$ between -10 to 10 \kms
to search for any significant CO disc emission that could be present
in the data. At each radius we estimate the uncertainty based on the
rms on each channel, the number of independent spectral resolution
elements considered
($\sim\tfrac{\mathrm{number\ of\ channels}}{2.667}$, due to Hanning
smoothing\footnote{https://safe.nrao.edu/wiki/pub/Main/ALMAWindowFunctions/\\Note\_on\_Spectral\_Response.pdf})
and the number of independent beams around each ellipse
($\sim\tfrac{\mathrm{Perimeter\ of\ ellipse}}{\mathrm{beam\ major\ axis}}$).

\begin{figure}
  \includegraphics[trim=0.0cm 0.0cm 0.0cm 0.0cm, clip=true,
    width=1.0\columnwidth]{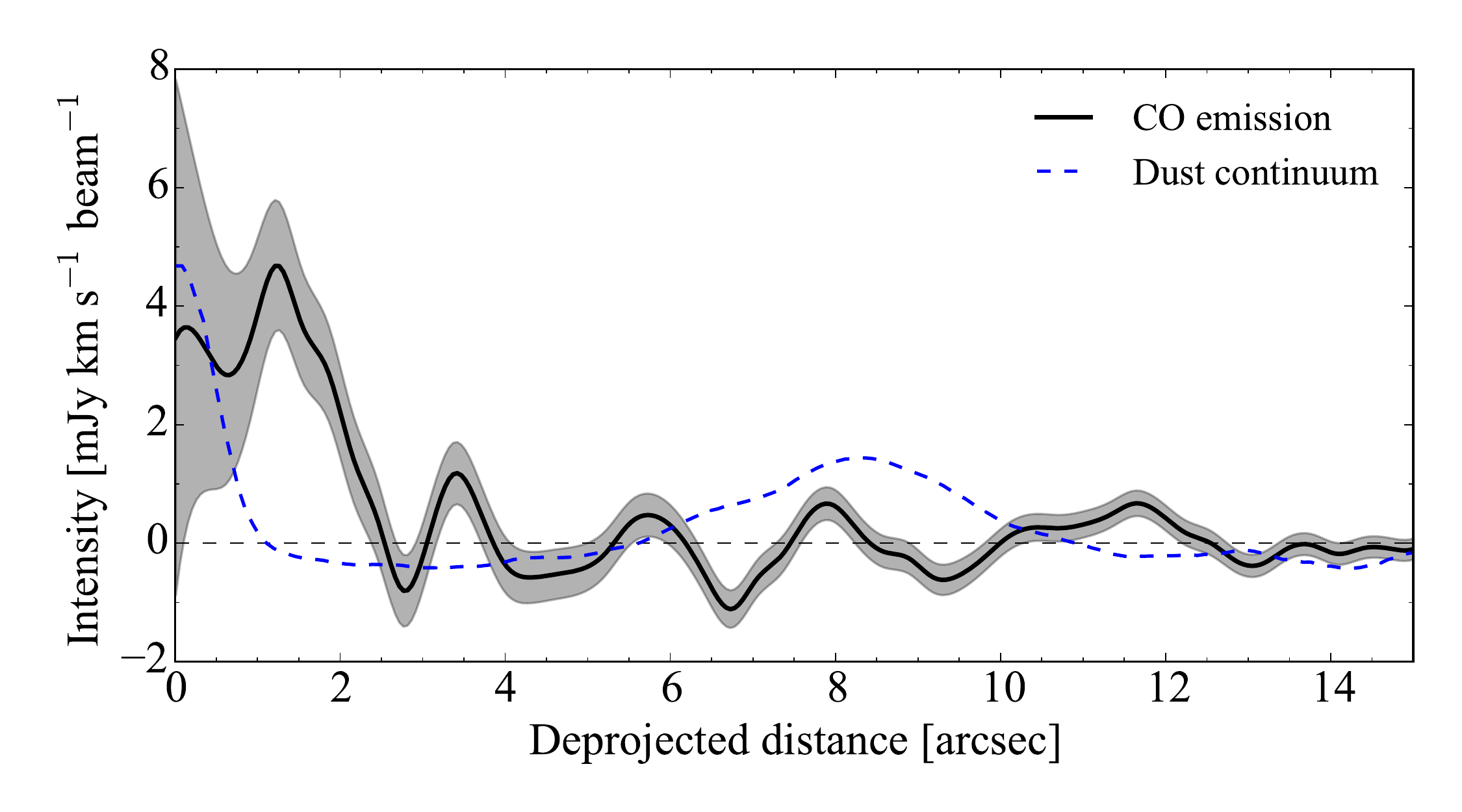}
  \caption{Mean CO intensity vs distance to the star obtained by
    azimuthally averaging the continuum subtracted dirty channel maps
    and integrating over the line width expected due to Keplerian
    rotation, assuming an heliocentric stellar radial velocity of 1.5
    km~s$^{-1}$. The grey areas represent the 68\% confidence region.}
    \label{fig:Irco}
\end{figure}

In Figure \ref{fig:Irco} we present the mean intensity profile
spectrally integrated assuming $v_\star=1.5$~km~s$^{-1}$. The
continuum radial profile is overlayed in blue dashed line. We find an
intensity peak of $4.7\pm1.1$ mJy~beam$^{-1}$~\kms at
$\sim1\farcs2$. This peak is present above $3\sigma$ for $v_\star$
ranging from -0.5 to 5.0 km~s$^{-1}$. Apart from the peak at
$\sim1\farcs2$ from the star, no significant emission is detected. The
peak of the emission corresponds to a radius of $22\pm6$~AU, where the
uncertainty is estimated as the beam semi-major axis divided by the
S/N. This implies that the putative detection is not strictly
co-located with the hot dust. Note that the sensitivity of the
intensity profile varies with radius because the number of beams and
spectral resolution elements over which we are integrating changes
with radius. Therefore, it is possible that similar or even higher CO
levels are present within $1\farcs2$, where the integrated sensitivity
is lower.
%%  However, CO emission could be present closer in below our
%% detection limit extending outwards.

\begin{figure}
  \includegraphics[trim=0.5cm 0.0cm 0.4cm 0.0cm, clip=true,
    width=1.0\columnwidth]{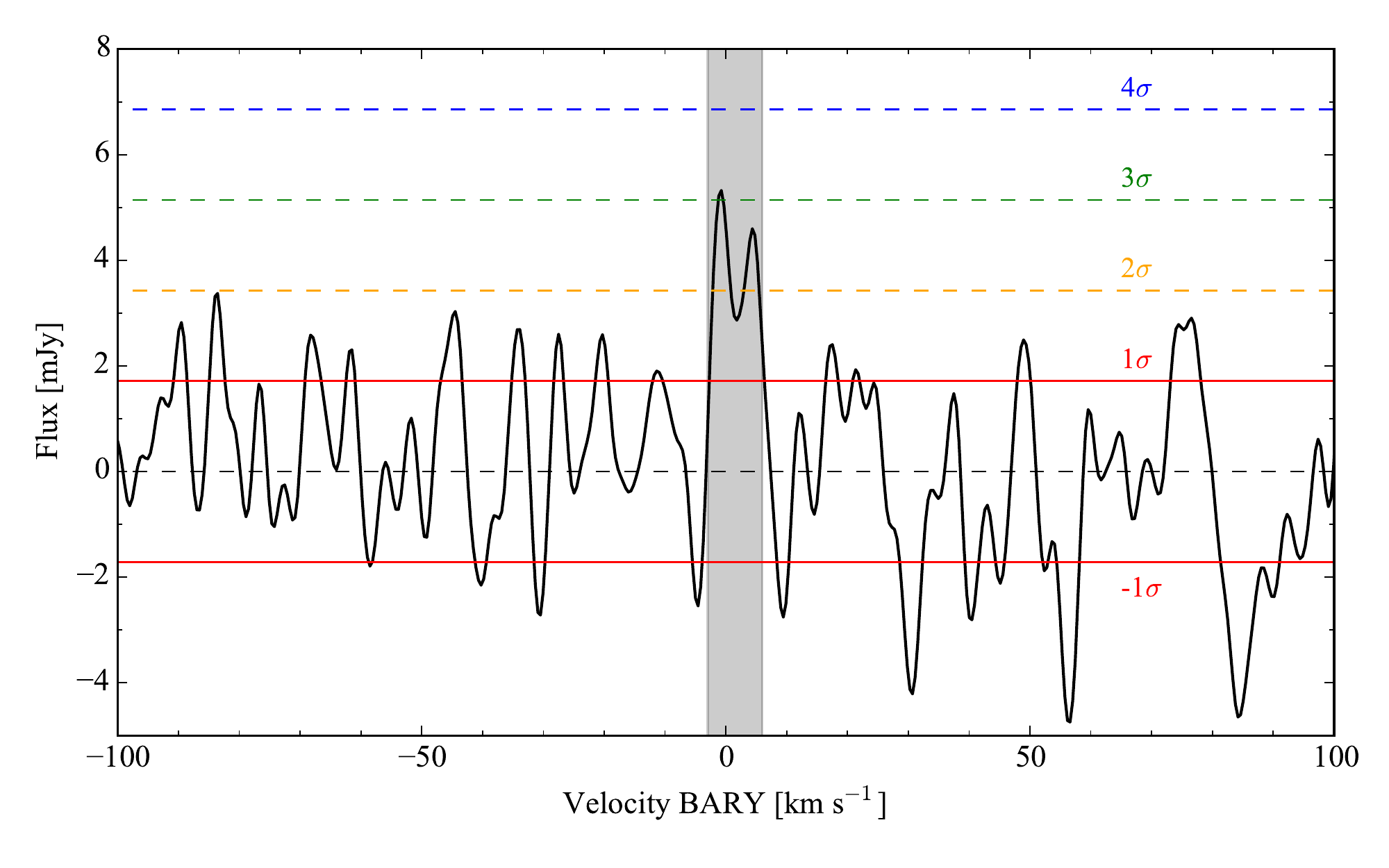}
  \caption{Continuum subtracted integrated spectrum inside an elliptic
    mask of minimum and maximum semi-major axis of $0\farcs8$ and
    $2\farcs0$, and oriented as the dust continuum outer belt. The
    original spectrum was smoothed with a Gaussian kernel with
    standard deviation of 2.5 channels. The horizontal lines represent
    $\pm1$, 2, 3 and 4$\sigma$ levels. The grey region represents
    velocities between -2.7 \kms and 6.3 \kms, equivalent to the line
    width expected due to Keplerian rotation at $\sim$20 AU
    ($1\arcsec$). The velocities represent the Doppler shift with
    respect to 345.796 GHz in the Barycentric reference frame.}
    \label{fig:cospectrum}
\end{figure}

In Figure \ref{fig:cospectrum} we present a spectrum obtained by
integrating the dirty map over an elliptic mask oriented as the outer
belt and with an inner and outer semi-major axes of $0\farcs8$ to
$2\farcs0$ (15 to 37 AU), respectively, to maximise the S/N. We also
smoothed the spectrum convolving it with a Gaussian kernel with a
standard deviation of 2.5 channels or 1.1 \kms. The spectrum shows a
double peaked line centered at 1.8, consistent with the expected
profile for a ring of CO gas rotating around the star at a radius of
$\sim20$~AU. The total line flux between -2.7 \kms and 6.3 \kms is
$38\pm9$~mJy~\kms ($4\sigma$). This is used in Sec. \ref{sec:modelco}
to estimate the mass of CO that could be present at this location. If
the gas is in Keplerian rotation, the emission should have a butterfly
pattern, with half of the emission having a positive or negative
Doppler shift and arise from either the south east or north west half
of the disc (separated by the minor axis of the disc). Thus, we try to
integrate the emission on one half of the disc for positive Doppler
shifts and on the other half for negative Doppler shifts to reduce the
integrated noise. We find similar S/N ratios for both possible disc
orientations, slightly preferring the north-western and south-eastern
halves having positive and negative Doppler shifts, respectively, but
not at a significant level. With the preferred orientation we obtain a
S/N of 3.9 for the integrated flux, while with the opposite the S/N is
3.4. The small difference in S/N shows that there is emission with the
same Doppler shift on both halves of the disc, in contradiction with
the Keplerian and disc orientation assumption. This could be due to
the emission being not very well constrained in position as we are
using the dirty channel maps, the uncertain stellar radial velocity,
the disc orientation being different compared to the outer belt, or
deviations from Keplerian rotation. We can also estimate the
likelihood of that any search of CO conducted over the range
$v_\star=\pm10$~km~s$^{-1}$ and over annuli with semi-major axes
within the range $0-8\arcsec$ would result in a $4\sigma$
detection. We find that this has a probability of 1\%. The origin of
this emission if real is discussed in Sec. \ref{dis:co}. Deeper ALMA
observations are necessary to confirm this detection and study the gas
rotation pattern.

In order to place upper limits on the total CO emission from the outer
belt and inner component (co-located with the hot dust), we compute
the uncertainty on the total flux given the noise level on each
channel map and the number of velocity channels to consider. Around
the outer belt we obtained an integrated noise level of
12~mJy~km~s$^{-1}$ integrating azimuthally and radially along the belt
seen in continuum emission (from 130 AU to 170 AU), and in velocity at
the expected radial velocity on each pixel due to Keplerian rotation
and $37^{\circ}$ inclination. For the inner component, we integrate
the emission inside $1\farcs2$ between $\pm21$ \kms (expected line
width assuming no emission is coming from a radii smaller than 1 AU)
obtaining an integrated noise level of 11 mJy~\kms.  Based on this, in
Sec. \ref{sec:modelco} we also calculate a $3\sigma$ upper limit on
the CO mass that could be in the outer belt and close in co-located
with the hot dust.

\section{Disc modelling}
\label{sec:model}

As the image reconstruction suffers from strong artefacts, we analyse
the ALMA observations in the uv space and infer the mm-sized dust
distribution by fitting the observed visibilities in the continuum
with four different debris disc models: 1) an outer belt with a
radial Gaussian surface density distribution to constrain the mean
radius and width of the belt; 2) a self-stirred disc following the
parametrization presented in \cite{Kennedy2010} to study if the disc
is consistent with such a scenario; 3) a LHB-like density
distribution, which connects the outer belt with the inner regions
with a surface density proportional to $r^{1.5}$, that then we relax
to a simple axisymmetric double power law distribution; 4) an
eccentric outer belt to constrain the disc global eccentricity.

The models consist of a central star modelled using a Kurucz template
spectrum\footnote{http://www.stsci.edu/hst/observatory/crds/k93models.html}
\citep{Kurucz1979} with an effective temperature of 7000 K and a
stellar radius of 1.75 $R_\odot$ to fit the stellar flux
($\sim250$~$\mu$Jy at 0.88 mm) as we impose the luminosity of the star
to be $4\pi R_\star^{2}\sigma T_\star^{4}$. The star is surrounded by
a dusty disc with grains formed by astrosilicates \citep{Draine2003},
amorphous carbon \citep{LiGreenberg1997} and water ice
\citep{LiGreenberg1998}, with mass fractions of 70\%, 15\% and 15\%,
respectively, with an internal density of 2.9 g~cm$^{-3}$ and mixed
using the Bruggeman's rule \citep{BohreHuffman1983} to match the
composition used in \cite{Duchene2014}. We assume a Dohnanyi-like
grain size distribution with a power law index of -3.5, with minimum
and maximum grain size of 1.0~$\mu$m and 1.0 cm, respectively. This
leads to a mass-weighted absorption opacity
$\kappa_{\mathrm{abs}}=3.8$~cm$^{2}$~g$^{-1}$ at 0.88 mm, computed
using the Mie theory code of \cite{BohreHuffman1983}. While
$(a_{\min}, a_{\max})$, $\kappa_{\mathrm{abs}}$ and the derived total
dust mass are highly dependent on our choice of the grain composition
and size distribution, these assumptions have very little effect on
the derived disc structure. Finally, we assume a Gaussian vertical
mass distribution parametrized by a scale height $\Hr$ that scales
linearly with radius as $\Hr=\mathrm{h}r$. As shown in
\cite{Marino2016}, this can be constrained by resolved ALMA
observations of axisymmetric discs; therefore we leave h as a free
parameter.

%% The star is modelled using a Kurucz template
%% spectrum\footnote{http://www.stsci.edu/hst/observatory/crds/k93models.html}
%% \citep{Kurucz1979} with an effective temperature of 7000 K and a
%% stellar radius, $R_\star$, that we leave as a free parameter to fit
%% the stellar flux. This is necessary because the absolute flux
%% uncertainty in the ALMA data is larger than the uncertainty on the
%% stellar flux at this wavelength. In our models we assume a Gaussian
%% vertical mass distribution parametrized by a scale height $\Hr$ that
%% scales linearly with radius as $\Hr=\mathrm{h}r$. As shown in
%% \cite{Marino2016}, this can be constrained by resolved ALMA
%% observations of axisymmetric discs; therefore we leave h as a free
%% parameter.
%% 1.7 $R_{\odot}$, which matches the photospheric emission in the
%% mid-infrared \citep[$\sim$250 $\mu$Jy,][]{Sylvester1996}.

Using
RADMC-3D\footnote{http://www.ita.uni-heidelberg.de/$\sim$dullemond/software/radmc-3d/}
\citep{RADMC3D0.39} we solve the thermal equilibrium of the mean dust
species defined above, obtaining a temperature field that varies with
radius as $42(r/150\ \mathrm{AU})^{-0.5}$~K, that then is used to
produce synthetic images of the system at 0.88 mm. Finally, models are
compared with the observations simulating model visibilities with the
same uv-coverage and pointing offsets.

%% CASA 4.4 as follows: with the
%% \textit{ft} we compute model visibilities based on our synthetic
%% images and using the same uv-sampling as the observations, and then
%% using \textit{uvsub} we subtract the model to the data.

A Bayesian approach is used to constrain the different parameters of
the disc models (details below), sampling the parameter space to
recover the posterior distribution with the public python module
\textit{emcee} that implements the Goodman \& Weare's Affine Invariant
MCMC Ensemble sampler \citep{GoodmanWeare2010, emcee}. Before
running our MCMC routine, we reduced the original ALMA data averaging
the visibilities with a time and frequency bins of 3 min and 2 GHz,
respectively. This reduces considerably the computational time of the
MCMC routine without losing significant information (checked by
comparing the S/N in the CLEAN images before and after averaging the
data) and without producing a bandwidth or time smearing bigger than
$0\farcs15$ along the ring. The posterior distribution is then defined
as the product of the likelihood function and the prior probability
distribution functions for each parameter, which we assume are
uniform. The likelihood function is defined proportional to
$\exp(-\chi^2/2)$, with $\chi^2$ defined as
\begin{equation}
  \chi^2=\sum_{i} \frac{||V_\mathrm{data, i}-V_\mathrm{model, i}||^{2}}{\delta V_\mathrm{data, i}^2},
\end{equation}
where the sum goes over the $1.16\times10^{6}$ uv points of the
observed visibilities, $V_\mathrm{data, i}$, previously averaged. The
estimated error $\delta V_\mathrm{data,i}$ is calculated based on the
intrinsic dispersion of the visibilities over one scan with the task
\textit{statwt} from CASA 4.4. In our priors, we impose a lower limit
to the vertical aspect ratio, h, equivalent to 0.03. This is necessary
as our model has a fixed grid with a vertical resolution of 2.3 AU at
150 AU ($h=$0.015) around the midplane, chosen to reduce the
computational time of our MCMC routine. We run emcee for 300-1000
steps using 50-180 \textit{walkers}, depending on the model, obtaining
auto-correlation lengths of $\lesssim50$, small enough to have more
than 300 independent sets of samples. The different models are
detailed below and compared in Sec. \ref{sec:modelcomparison} using
the Bayesian information criterion \citep[BIC,][]{Schwarz1978}.

%% This value is consistent with the minimum h expected in the absence
%% of perturbing planets \citep{Thebault2009}.

\subsection{A radially symmetric belt}
\label{sec:modelring}

In order to estimate the mean radius and radial width of the outer
belt, our first model consists of a belt with a surface density
distribution that is parametrized with a Gaussian centered at $r_0$
and with a radial full width half maximum (FWHM) $\Delta r$ as
\begin{equation}
  \rho(r,z)=\rho_0 \exp\left[ -\frac{4 \ln(2) (r-r_0)^2}{\Delta r^2}
    -\frac{z^2}{2 \Hr^{2}} \right], \label{eq:gausring}
\end{equation}
where $\rho_0$ is defined to match the total dust mass, $M_d$, and
$\Hr=\mathrm{h}r$ is the scale height. Then, the dust density
distribution is defined by $M_d$, $r_0$, $\Delta r$ and $h$ that we
leave as free parameters. Two extra free parameters were defined to
fit the disc inclination, $i$, and position angle, PA, in sky. In
Table \ref{table:gaussian} we summarise the best fit values and
uncertainties. We find the disc mean radius is $152\pm3$ AU and a
radial FWHM of $46\pm5$ AU, consistent with our previous estimate in
Sec. \ref{sec:continuum}.

%% In Figure \ref{fig:corner_gaussian} we present the posterior
%% distributions of $\Delta r$, h and $r_0$, while .  

%% \begin{figure} \includegraphics[trim=0.5cm 0.5cm 0.0cm 0.0cm,
%%   clip=true,
%%   width=1.0\columnwidth]{corner_drhr_gauss.pdf} \caption{Posterior
%%   distribution of h$=\Hr/r$, $r_0$ and $\Delta r$ from the belt
%%   model. The vertical dashed lines represent the 16th, 50th and
%%   84th percentiles. Contours correspond to 68\%, 95\% and 99.7\%
%%   confidence regions. This plot was generated using the python
%%   module \textit{corner}
%%   \citep{cornerplot}. } \label{fig:corner_gaussian} \end{figure}

\begin{table}
  \centering
  \caption{Belt model best fit values. Median $\pm$ uncertainty based
    on the 16th and 84th percentile of the marginalised
    distributions. For parameters with distributions extending out to
    the minimum or maximum allowed values, a $1\sigma$ upper or lower
    limit is specified based on the 68th or 32nd percentile,
    respectively.}
  \label{table:gaussian}
  \begin{tabular}{lc} %cc} % 7 columns, r,dr,h,pa,inc,raoff,decoff 
    \hline
    \hline
    Parameter & Best fit value \\
    \hline
    %% $R_\star$ [$R_\odot$] & $1.46\pm0.08$ \\
    $M_{d}$ [M$_{\oplus}$] & $0.014\pm0.001$  \\
    $r_{0}$ [AU] & $152\pm3$ \\
    $\Delta r$ [AU] &  $46\pm5$  \\
    $h$ &  $<0.13$ ($1\sigma$)  \\ %$0.10^{+0.05}_{-0.04}$  \\
    PA [$^{\circ}$] &  $117\pm4$ \\
    $i$ [$^{\circ}$] & $35\pm2$ \\
    %% RA offset [$\arcsec$] & $-0.03\pm0.08$  \\
    %% Dec offset [$\arcsec$] &  $0.00\pm0.07$ \\
    \hline
  \end{tabular}
\end{table}

We derive an inclination and PA of $35\pm2^{\circ}$ and
$117\pm4^{\circ}$, respectively, that match the previous estimations
by \cite{Lebreton2016} by fitting ellipses to \textit{Herschel}
observations ($i=38.2\pm3.6^{\circ}$ and PA$=116.2\pm1.1^{\circ}$),
but our derived inclination is significantly smaller than
$47\pm1^{\circ}$, the inclination derived by \cite{Duchene2014} using
the same \textit{Herschel} data, but fitting a disc model to the
observations and using a different data reduction. This difference
could be due to the different methods used to derived the disc
inclination, data reduction, or due to different PSFs as these can
vary significantly between different \textit{Herschel} observations
\citep[e.g. ][]{Kennedy2012}, making their uncertainties larger than
the ones quoted above. The inferred disc mean radius and radial width
are overall consistent with their derived dust distribution, with
differences due to the different models assumed. The total flux of the
best fit model is $13.3\pm1.0$ mJy (excluding calibration
uncertainties). This is higher than the measured flux from the CLEAN
image ($9.2\pm0.5$~mJy) and consistent with the total flux measured by
SCUBA-2/JCMT. We also find that the posterior distribution of the
aspect ratio $h$ peaks at 0.10, $\sim2\sigma$ above the lower limit in
our prior distribution (0.03), but still consistent with 0.

%% Finally, we find that the RA and dec offsets are smaller than the
%% beam size and consistent with zero within the estimated
%% uncertainties and ALMA astrometric accuracy
%% ($\sim0.1\arcsec$\footnote{https://help.almascience.org/index.php?/Knowledgebase/Article/View/319/6/what-is-the-astrometric-accuracy-of-alma}).

\begin{figure*}
  \includegraphics[trim=0.0cm 0.5cm 0.0cm 0.5cm, clip=true,
    width=1.0\textwidth]{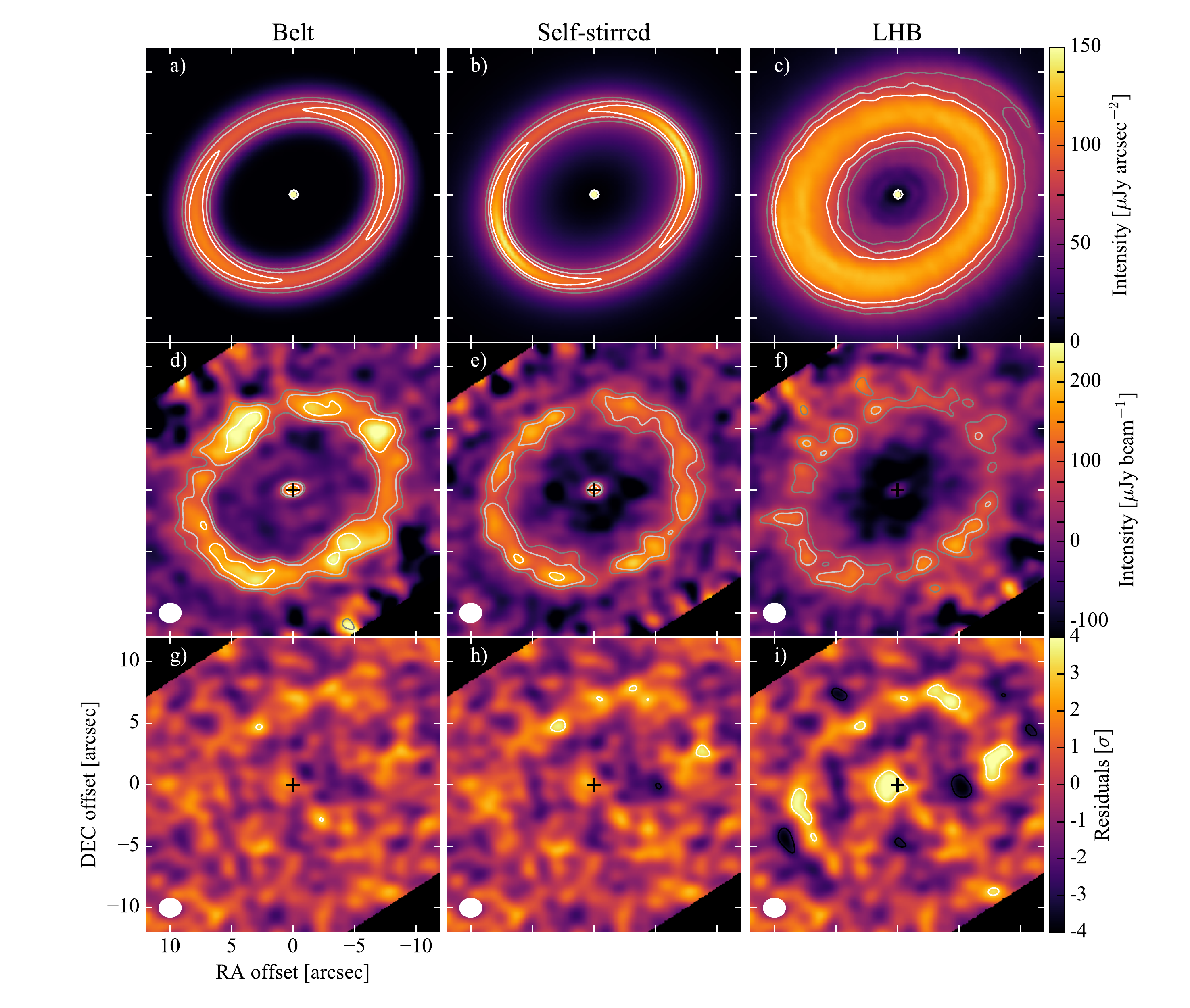}
  \caption{Simulated images at 0.88 mm. The first, second and third
    column correspond to the best-fit belt model
    (Sec. \ref{sec:modelring}), best-fit self-stirred disc model
    (Sec. \ref{sec:modelselfstirred}) and Late Heavy Bombardment model
    (Sec. \ref{sec:modellhb}), respectively. First row: synthetic
    images of the disc. Contours represent 65, 80 and 95
    $\mu$Jy~arcsec$^{-2}$. Second row: primary beam corrected
    simulated CLEAN images using the same uv-sampling, adding Gaussian
    noise to the visibilities according to the variance of the
    observations, and uv-tapering the visibilities with a Gaussian of
    FWHM of $1\farcs2$ in sky. Contours represent 2, 3 and 5 times the
    local noise level. Third row: Dirty map of the residuals after
    subtracting the model visibilities from the ALMA observations. The
    noise level on the residuals is uniform and equal to
    30~$\mu$Jy~beam$^{-1}$ as they are not corrected by the primary
    beam. The black and white contours represent $\pm3\sigma$. The
    beam size is represented by a white ellipse in the bottom left
    corner. The x- and y-axes indicate the offset from the stellar
    position in R.A. and decl. in arcsec, i.e. north is up and east is
    left. The stellar position is marked with a black
    ``+''. \label{fig:models}}
\end{figure*}

%% The marginalised posterior
%% distributions is presented in Figure \ref{fig:corner_ecc}.

%% \begin{figure}
%%   \includegraphics[trim=0.0cm 0.0cm 0.0cm 0.0cm, clip=true,
%%     width=1.0\columnwidth]{corner_rdrew.pdf}
%%   \caption{Posterior distribution of $\Delta r$, $r_0$, $e$ and
%%     $w$. The vertical dashed lines represent the 2.5th, 50th and
%%     97.5th percentiles. Contours correspond to 50\%, 75\% and 90\%
%%     confidence regions. This plot was generated using the python
%%     module \textit{corner} \citep{cornerplot}. }
%%     \label{fig:corner_ecc}
%% \end{figure}

%% Moreover, to study differences between the inner and outer edge of the
%% disc, we fit a model with two different Gaussian radial distributions
%% with different $\Delta r$'s for $r<r_0$ and $r>r_0$, but matched in
%% density at $r_0$. The surface density is defined as
%% \begin{equation}
%%  \rho(r,z) = \begin{cases}
%% \rho_0 \exp\left[ -\frac{4 \ln(2) (r-r_0)^2}{\Delta r_\mathrm{in}^2}
%%     -\frac{z^2}{2 \Hr^{2}} \right] & \text{$r<r_0$}\\
%% \rho_0 \exp\left[ -\frac{4 \ln(2) (r-r_0)^2}{2\Delta r_\mathrm{out}^2}
%%     -\frac{z^2}{2 \Hr^{2}} \right] & \text{$r>r_0$}.
%% \end{cases}
%% \end{equation}
%% We find that the observations are best fitted with $\Delta
%% r_\mathrm{in}=46\pm13$ and $\Delta r_\mathrm{out}=49\pm15$, both in
%% agreement with $\Delta r$ of our first model described before. That
%% is, there is no evidence that the inner and outer edges of the belt
%% have a profile that is asymmetric about the midpoint of the belt.

A synthetic image, simulated ALMA image, and dirty map of the
residuals of the best fit model are presented in Figure
\ref{fig:models}a, d and g, respectively. Despite the model being
axisymmetric, the simulated ALMA image shows asymmetric structure
along the belt, similar to the observed features in the ALMA image
(see Fig. \ref{fig:cleannatural}). The maxima along the minor axis and
minima along the major axis are well reproduced; thus, the observed
asymmetries are likely an artefact of the image reconstruction caused
by a low S/N in the visibilities and the uv-sampling. After
subtracting the model visibilities, the dirty map of the residuals
shows no excess above $3\sigma$ that could be attributed to an
overdensity in the belt or emission inside the cavity.

\begin{figure}
  \includegraphics[trim=0.0cm 0.0cm 0.0cm 0.0cm, clip=true,
    width=1.0\columnwidth]{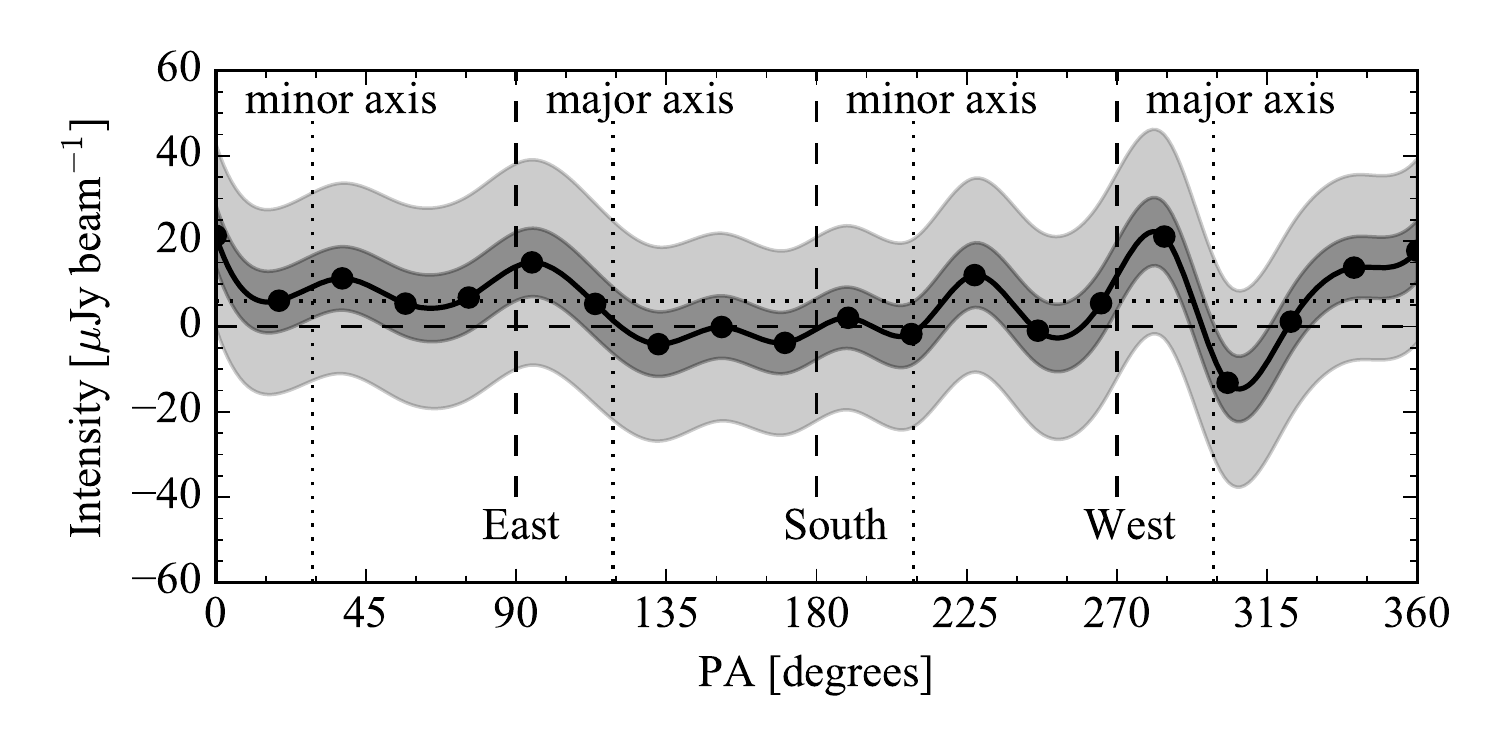}
  \caption{Intensity azimuthal profiles of the dirty map of the
    residuals of the best fit belt model, obtained averaging over
    elliptic annuli of semi-major axes between $6\farcs8$ and
    $9\farcs8$ (equivalent to 125 and 180 AU) and over 18$^{\circ}$
    wide wedges (centered at the black dots). The horizontal dotted
    line represents the mean intensity around the residuals in the
    outer belt. The light and dark grey areas represent the 68\% and
    99.7\% confidence region.}
    \label{fig:Iphires}
\end{figure}

Similar to the analysis in Sec. \ref{sec:continuum}, in Figure
\ref{fig:Iphires} we compute the azimuthal profile of the residuals
along the outer belt averaging between 130 and 175 AU. The integrated
flux of the residuals around the outer belt is $1.2\pm0.3$ mJy
(correcting by the primary beam), an excess which is within the
uncertainty of the flux in our model. At a PA of $290^{\circ}$ the
residuals show an excess of 20~$\mu$Jy~beam$^{-1}$
($3\sigma$). Moreover, the excess on the north-west half of the disc
is $0.8\pm0.2$ mJy, while $0.4\pm0.2$ mJy on the opposite side of the
disc. This is consistent with what we found in
Sec. \ref{sec:continuum} and the 70 $\mu$m PACS image of the disc, and
points to a slight excess of emission in the north-west side of the
disc.

%% This is marginal evidence for a slight emission excess on the
%% north-west side of the disc where Herschel images also showed
%% excess emission \citep{Duchene2014}.  No residuals larger than
%% $3\sigma$ are present inside the cavity.

The best fit model has a reduced chi-squared
$\chi^2_\mathrm{red}=1.0037898$, with $6.54\times10^6$ different
visibility measurements before averaging. This is used later to
compare the goodness of fit of different models.

%% \textcolor{red}{It is worth
%%   noting that the peak intensity that we found in
%%   Sec. \ref{sec:continuum} at a PA of 230$^{\circ}$ is not present in
%%   the residuals, thus it is likely to be an artefact from the image
%%   deconvolution as the simulated ALMA image shows (see Figure
%%   \ref{fig:models}d). On the other hand, inside the cavity, a
%%   $3\sigma$ peak appear to the north-east separated by $\sim0\farcs4$
%%   from the stellar position. Moreover, $\sim3\sigma$ residuals are
%%   also present within $2\arcsec$ at the East of the stellar
%%   position. However, inside the cavity we also find $\sim3\sigma$
%%   negative residuals. Therefore, it is unlikely that the observed
%%   positive residuals are real.}

\subsection{Self-stirred disc}
\label{sec:modelselfstirred}

The cavity and outer belt and in \etacorvi \ could be the result of a
primordial depletion of solids from a few AUs up to 110 AU, followed
by an overdensity or belt of planetesimals with a peak at 150 AU, like
in a transitional protoplanetary disc
\citep[e.g.,][]{Espaillat2014}. Alternatively, the current disc
structure could just be the result of the collisional evolution of a
broad primordial disc of planetesimals being self-stirred as
Pluto-sized objects are being formed at different epochs at different
radii. In order to test the latter scenario and see if the disc
morphology can be explained by self-stirring, we considered a second
model that consists of a parametric self-stirred disc based on the
work by \cite{Kennedy2010}. In this scenario, the surface density of
solids evolves from a primordial disc of planetesimals to a stirred
disc in collisional equilibrium as
\begin{equation}
 \Sigma(r,t) = \begin{cases}
   \Sigma(r,0) \ x_\mathrm{delay}  & \text{$t<t_\mathrm{stir}(r)$}
   \\
   \Sigma(r,0)/\left[1+\frac{t-t_\mathrm{stir}(r)}{t_c(r)}\right]  & \text{$t>t_\mathrm{stir}(r)$}
\end{cases} \label{eq:self-stir}
\end{equation}
where $\Sigma(r,0)$ is the primordial surface density of solids
assumed here to follow a power law, i.e. $\Sigma_0(r/r_0)^{-\gamma}$,
and $t$ is the age of the system (assumed to be 1.4 Gyr here). At each
radius the disc is stirred at epoch $t_\mathrm{stir}(r)$ after which
it is assumed to be in collisional equilibrium, with a collisional
lifetime of the biggest planetesimals $t_c(r)$ at the initial epoch of
the collisional cascade (i.e. after being stirred). Similar to
\cite{Kennedy2010}, we also introduce a factor $x_\mathrm{delay}\leq1$
in Eq.  \ref{eq:self-stir} when $t<t_\mathrm{stir}$ because the
collisional cascade has not yet begun, thus the amount of mass in
grains that contribute to the millimetre emission could be lower
(different size distributions).

The timescale at which Pluto-sized objects are formed depends both on
the distance to the star and the surface density of solids at that
distance. Generally speaking, this timescale will be proportional to
the orbital period and inversely proportional to the primordial
surface density of solids \citep[e.g.][]{Lissauer1987},
i.e. $t_\mathrm{stir}(r)\propto r^{3/2}/\Sigma(r,0)$. Because we aim
to fit the observations, in our self-stirred model we use the
following empirical relation
\begin{equation}
  t_\mathrm{stir}(r)=t_\mathrm{age}(r/r_\mathrm{stir})^{3/2+\gamma}, \label{eq:tstir}
\end{equation}
where $r_\mathrm{stir}$ is defined as a reference radius where
stirring is happening at the present epoch ($t_\mathrm{age}$). With
this parametrization the peak of the disc emission is at
$r_\mathrm{stir}$ which is comparable to $r_{0}$ in the belt model. We
can make an order of magnitude estimation of this timescale
extrapolating the timescale at which Pluto formed \citep[$\sim40$ Myr,
  see][and references therein]{Brown2002} to 150 AU using
$\gamma=1$. We find $t_\mathrm{stir}(150 \mathrm{AU})=1.5$ Gyr,
consistent with simulations by \cite{Kenyon2008} (see their Eq. 41),
and of the same order as the age of the system; therefore, we consider
that the self-stirring scenario is plausible.

On the other hand, the collisional lifetime of the biggest
planetesimals in the disc is \citep[see
][]{Wyatt2007collisionalcascade}
\begin{equation}
  t_c(r) \ \propto \ r^{7/3} \ D_c \ Q_\mathrm{D}^{\star 5/6} \ \Sigma(r,0)^{-1},
\end{equation}
where $D_c$ is diameter of the biggest planetesimal, and $\Qd$ is the
disruption threshold of planetesimals, here assumed to be independent
of size. The expression above can be rewritten as
\begin{equation}
  t_c(r)=t_0 (r/r_\mathrm{stir})^{7/3+\gamma},
\end{equation}
where $t_0$ is defined as the collisional lifetime of the biggest
objects at $r_\mathrm{stir}$. The net result is that the surface
density is proportional to $r^{7/3}$ when $r\ll r_\mathrm{stir}$ and
$r^{-\gamma}$ when $r>r_\mathrm{stir}$, while its shape close to
$r_\mathrm{stir}$ is determined by the ratio $t_c/t_\mathrm{stir}$
which is $\approx t_0/t_\mathrm{age}$.

\begin{figure}
  \includegraphics[trim=0.0cm 0.0cm 0.0cm 0.0cm, clip=true,
    width=1.0\columnwidth]{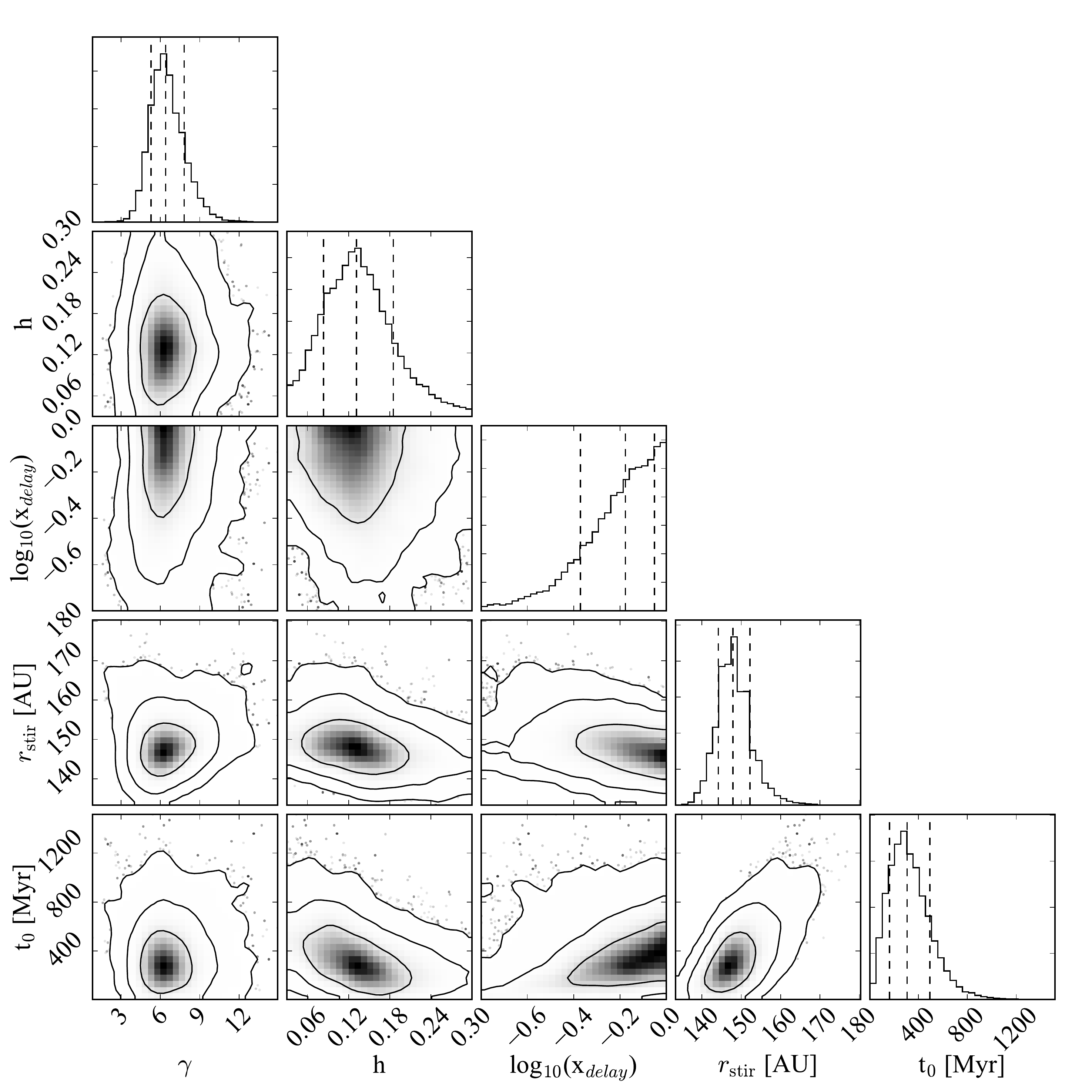}
  \caption{Posterior distribution of $h=\Hr/r$,
    $\log_{10}(x_\mathrm{delay})$, $r_\mathrm{stir}$ and $t_0$ from
    the self-stirred model.  The vertical dashed lines represent the
    16th, 50th and 84th percentiles. Contours correspond to 68\%, 95\%
    and 99.7\% confidence regions. This plot was generated using the
    python module \textit{corner} \citep{cornerplot}. }
    \label{fig:corner_selfstir}
\end{figure}

 Then, the free parameters are the total mass of dust $M_d$, h,
 $\gamma$, $x_\mathrm{delay}$, $r_\mathrm{stir}$, $t_0$, PA and
 $i$. The posterior distributions of the most relevant parameters are
 presented in Figure \ref{fig:corner_selfstir} and in table
 \ref{table:selfstir} we summarise the best fit parameter values and
 uncertainties. The total flux of the best fit model is $16.7\pm2.3$
 mJy. From the best fit values of the self-stirred model, we find that
 $\gamma=6.4\pm1.4$, $x_\mathrm{delay}=0.3-1.0$ (95.0\% confidence,
 with a marginalised posterior distribution that peaks at 1), and
 $t_0<t_\mathrm{age}$, i.e. $t_c<t_\mathrm{stir}$ (99.7\%
 confidence). This means that the disc has a sharp inner and outer
 edge (see Figure 2 in \cite{Kennedy2010}). The disc model image,
 simulated ALMA image and dirty map of the residuals are presented in
 Figure \ref{fig:models}b, e and h, respectively.  Negative and
 positive emission at $3\sigma$ levels appear close to the stellar
 position in the simulated ALMA image and dirty map of the residuals,
 respectively. However, these negatives are less significant in the
 belt model, which does not have any emission in the inner regions
 apart from the star (see Sec. \ref{sec:modelring}). If the emission
 were really there, then it should appear brighter in the residuals of
 the belt model. We find the best fit model has a
 $\chi^2_\mathrm{red}=1.0037958$, with $6.54\times10^6$ different
 visibility measurements.

 %% Therefore, we conclude that these could be caused by the low S/N,
 %% $u-v$ sampling of the ALMA data and the dirty map deconvolution
 %% algorithm.

 %% The residuals close to the stellar position seem to be
 %% artefacts as the ring model, which does not have any emission in the
 %% inner regions apart from the stellar, does not show any significant
 %% excess at those locations. Therefore, we conclude these are caused by
 %% the low S/N, $u-v$ sampling of the ALMA data and the dirty map
 %% deconvolution algorithm.
 
 %% \textbf{Discuss residuals on ring and self-stirred sections}
 
%% No significant emission ($\gtrsim 3$) is found in the residuals

\begin{table}
  \centering
  \caption{Self-stirred model best fit values. Median $\pm$
    uncertainty based on the 16th and 84th percentile of the
    marginalised distributions. For parameters with distributions
    extending out to the minimum or maximum allowed values, a
    $1\sigma$ upper or lower limit is specified based on the 68th or
    32nd percentile, respectively.}
  \label{table:selfstir}
  \begin{tabular}{lc} %cc} % 5 columns, r,dr,h,pa,inc,raoff,decoff 
    \hline
    \hline
    %% $R_\star$ [$R_\odot$] & $1.55\pm0.10$ \\
    $M_{d}$ [M$_{\oplus}$] & $0.019\pm0.003$\\
    $h$ & $<0.16$ ($1\sigma$) \\ %$0.13\pm0.05$ \\
    $\log_{10}(x_\mathrm{delay})$  & $>-0.26$ ($1\sigma$) \\ %$-0.2^{+0.1}_{-0.2}$ \\
    $r_\mathrm{stir}$ [AU]  & $148\pm4$  \\
    $t_0$ [Myr]& $300^{+180}_{-140}$  \\
    $\gamma$ & $6.4\pm1.4$ \\
    PA [$^{\circ}$] &  $118\pm4$ \\
    $i$ [$^{\circ}$] & $34\pm3$ \\
    %% RA offset [$\arcsec$] & $-0.04\pm0.09$  \\
    %% Dec offset [$\arcsec$] &  $0.01\pm0.07$ \\
    %% \\%& $0.482\pm6\times10^{-3}$\\ %&
    \hline
  \end{tabular}
\end{table}

Using Eq. 16 from \cite{Wyatt2008} and the derived values of
$r_\mathrm{stir}$ and $t_{0}$ we can estimate the mean surface density
($\Sigma=M_\mathrm{tot}/2\pi rdr$) or total mass in planetesimals
($M_\mathrm{tot}$), assuming a maximum planetesimal size
$D_\mathrm{c}$ of 1000 km and a power law size distribution of solids
with an exponent of -3.5. We find
\begin{equation}
  \begin{split}
  \Sigma_{r_\mathrm{stir},t=0}=0.7
  \left(\tfrac{r_\mathrm{stir}}{150 \ \mathrm{AU}}\right)^{7/3}
  \left(\tfrac{D_\mathrm{c}}{1000\ \mathrm{km}}\right)
  \left(\tfrac{\Qd}{200 \mathrm{J~Kg^{-1}}}\right)^{5/6}
  e^{-5/3}\\
  \left(\tfrac{M_\star}{1.4 \ \mathrm{M}_\odot}\right)^{-4/3}
  \left(\tfrac{t_0}{300\ \mathrm{Myr}}\right)^{-1}, \label{eq:sigmaprim}
  \end{split}
\end{equation}
in M$_{\oplus}$~AU$^{-2}$. This corresponds to a total mass of
$3\times10^{4}$ M$_{\oplus}$ in the outer belt, which is too massive
to be consistent with the self-stirred scenario. Considering only the
mass on solids, this is equivalent to a Toomre $Q$ parameter
\citep{Toomre1964} of 1 assuming a vertical aspect ratio of 0.1;
however, if we include the gas mass present during the protoplanetary
disc phase ($\Sigma_\mathrm{gas}\approx 100 \Sigma_\mathrm{solids}$)
we find that the disc would have been highly unstable under
gravitational perturbations. The high value of $M_\mathrm{tot}$ is due
to $t_c$ at $r_\mathrm{stir}$, i.e. $t_{0}$, which is too short for
1000 km sized bodies. This is because the model fits better the data
if the outer belt is narrow, i.e. $t_0\lesssim t_\mathrm{age}$.

Combining the mass derived above as a function of the maximum
planetesimal size, i.e. $M_\mathrm{solids}=3\times10^4
(Dc/1000\ \mathrm{km})$~M$_{\oplus}$, with dust mass on grains smaller
than 1 cm, i.e. $M_\mathrm{solids}=0.02
(Dc/1\ \mathrm{cm})$~M$_{\oplus}$, we find a maximum planetesimal size
of $D_\mathrm{c}\sim40$~m, but this contradicts the hypothesis that
the disc is stirred by the growth of planetesimals up to Pluto-sized
bodies. The collisional timescale could be longer (and so allow a
larger maximum planetesimal size) if the age of the system is greater
than 1.4 Gyr as it is the ratio between both that determines the shape
of the surface density; however, to reconcile the maximum planetesimal
size ($\sim1000$~km) and the disc mass, the age of the system should
be $\sim100$ times longer, which is impossible.  Therefore, we
conclude that self-stirring is unlikely to have produced the observed
disc morphology in \etacorvi. However, we cannot rule out that all the
solids are smaller than 40~m (leading to a total mass of $\sim1.3$
M$_\oplus$) if the disc is externally stirred, a possibility that is
discussed in Sec. \ref{dis:selfstir}.

\subsection{Late Heavy Bombardment}
\label{sec:modellhb}

As the hot dust around \etacorvi \ cannot be sustained by a
collisional cascade in situ, \cite{Lisse2012} suggested that this
system could be undergoing an instability similar to the Late heavy
bombardment (LHB) in the Solar System. The outward migration of
Neptune after Jupiter and Saturn crossed the 1:2 resonance caused a
massive delivery of planetesimals to the inner Solar System, high
frequency collisions with the Earth and Moon, and a population of
eccentric and highly inclined objects in the Kuiper Belt
\citep{Gomes2005}. During such instability the disc could have
specific observable features different from a scenario in which
planets are in stable orbits for long timescales. For example,
\cite{Booth2009} showed that the Kuiper belt should be broad during
the LHB (see Figure 1 therein) with a surface density profile
approximated by a power law proportional to $r^{1.5}$ between 1 and 27
AU and $r^{-4.8}$ between 38 and 106 AU. Moreover, the disc could
display asymmetries such as spiral arms or disc offsets with respect
to the stellar position produced by a planet put on an eccentric orbit
after the instability \citep[see Figure 7 in ][]{Pearce2014}.

\begin{figure}
  \includegraphics[trim=1.0cm 0.0cm 0.5cm 0.0cm, clip=true,
    width=1.0\columnwidth]{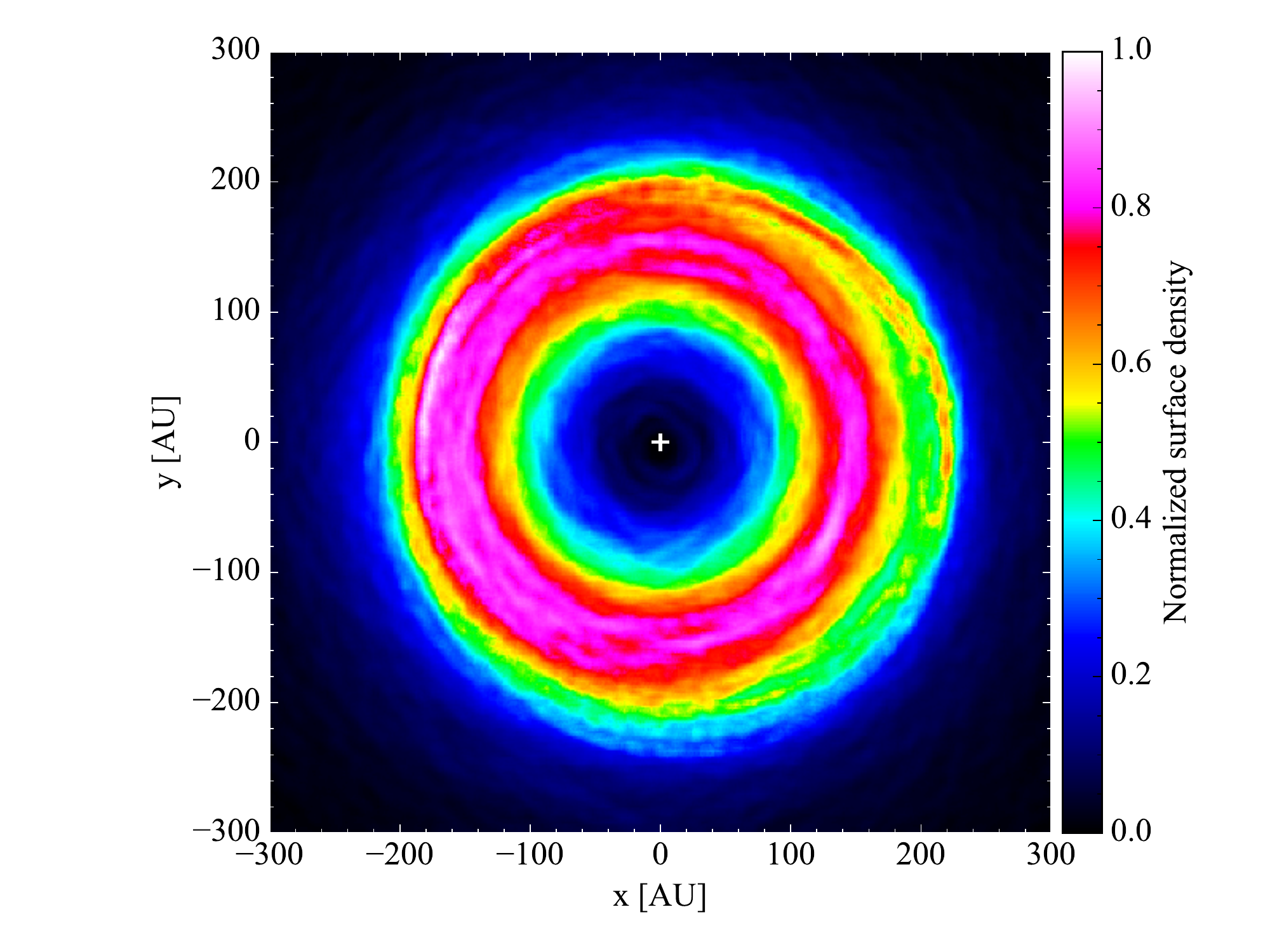}
  \caption{Normalized surface density of particles during the LHB
    computed from N-body simulations of one of the Nice models runs
    \citep{Gomes2005}. The particles from the simulation were
    duplicated assigning to each one a random anomaly in order to
    spread them around their orbits. Finally the particle's semi-major
    axes were scaled by 3.75 to match the size of the outer disc in
    \etacorvi. The stellar position is marked with a white ``+''.}
    \label{fig:lhb}
\end{figure}

Figure \ref{fig:lhb} shows the surface density of planetesimals during
the LHB in the Solar System (at $t=880$ Myr after the start of the
simulation), obtained from N-body simulations from one of the Nice
model runs \citep{Gomes2005}. As a first approximation we linearly
scale the semi-major axis of the planetesimals orbits to match the
location of the mean radius of \etacorvi, which maintains unchanged
the surface density proportional to $r^{1.5}$. This is not necessarily
true given the large size ratio between the Kuiper belt and the
\etacorvi \ debris disc (40/150) and the different ways the timescales
involved depend on the semi-major axis of particles in the
simulation. Therefore, N-body simulations tailored to \etacorvi
\ might be necessary; however, this first approximation can be used to
compare the observations with an LHB-like event in which an outer belt
is perturbed after a dynamical instability in the system, producing a
shallow surface density distribution of $r^{1.5}$ extending from the
outer belt down to the inner most regions (see Figure
\ref{fig:allmodels}). To smooth the derived surface density, we also
randomly spread each particle around its orbit according to the
velocity they have at each location. This is only valid if the
collisional timescales of the solids are longer than the orbital
period, which is probably true for mm-sized grains given the low
surface density of the disc. This technique also washes out any
resonant structure that relies on the correlation between the mean
longitude and the longitude of pericentre of a particle, although
resonances may not be a strong feature at this stage of the
evolution. Finally, we simulate images of the disc assuming that
mm-sized grains, which dominate the continuum emission at millimetre
wavelengths, have the same distribution as their parent bodies, as
they are not affected by radiation pressure. We vary the disc flux to
fit to the visibilities finding a best fit with a total disc flux of
$39\pm2$~mJy ($0.048\pm0.002$~M$_\oplus$ of dust, assuming the same
optical properties as the models described above). This is much higher
and inconsistent with the total flux measured by SCUBA-2/JCMT and is
required to fit the ALMA data because most of the emission is on large
scales, and thus, resolved out. In figure \ref{fig:models} c, f and i
we present the synthetic continuum image at 0.88 mm, ALMA simulation
and dirty map of the residuals after subtracting it to the observed
visibilities. Despite the fact that the model surface density displays
asymmetric features such as a spiral arms, we find that the simulated
observations are not sensitive enough to recover this structure, and
thus, they are roughly consistent with an axisymmetric disc. However,
the LHB best fit model fails to reproduce the observed surface
brightness observed by ALMA as it is broader and fainter in the
reconstructed image, despite requiring three times the flux measured
by SCUBA-2/JCMT.
%% This is also illustrated in Figure \ref{fig:res_profile}, where the
%% radial profile of the residuals shows significant positive and
%% negative emission.

In order to place better constraints on the surface density exponent
of a possible extended disc, we fit an axisymmetric model with a
surface density defined as a double power law
\begin{equation}
 \Sigma(r) = \begin{cases}
   \Sigma_0 \left(r/r_c\right)^{\alpha_\mathrm{in}} & \text{$r<r_c$}\\
   \Sigma_0 \left(r/r_c\right)^{\alpha_\mathrm{out}} & \text{$r>r_c$},
\end{cases}
\end{equation}
where $r_c$, $\alpha_\mathrm{in}$ and $\alpha_\mathrm{out}$ are free
parameters, together with the total mass of dust in the disc, $M_d$,
the vertical aspect ratio, h, and the PA and inclination of the disc
in sky. In Table \ref{table:2plaw} we summarise the best fit values
and uncertainties. We find $\alpha_\mathrm{in}=6.2^{+2.0}_{-1.3}$,
with a $3\sigma$ lower limit of 3.3. Therefore, we conclude that the
surface density of solids in the disc has to rise considerably
steeper than $r^{1.5}$ in the cavity towards the outer belt,
discarding a LHB-like surface density and a highly scattered outer
belt. The best fit model has a $\chi^2_\mathrm{red}=1.0037906$,
with $6.54\times10^6$ different visibility measurements.

\begin{table}
  \centering
  \caption{Double power law model best fit values. Median $\pm$
    uncertainty based on the 16th and 84th percentile of the
    marginalised distributions. For parameters with distributions
    extending out to the minimum or maximum allowed values, a
    $1\sigma$ upper or lower limit is specified based on the 68th or
    32nd percentile, respectively.}
  \label{table:2plaw}
  \begin{tabular}{lc} %cc} % 5 columns, r,dr,h,pa,inc,raoff,decoff 
    \hline
    \hline
    %% $R_\star$ [$R_\odot$] & $1.55\pm0.10$ \\
    $M_{d}$ [M$_{\oplus}$] & $0.016\pm0.002$\\
    $h$ & $<0.15$ ($1\sigma$) \\%$0.12^{+0.07}_{-0.04}$ \\
    $r_c$ [AU] & $151\pm4$\\
    $\alpha_\mathrm{in}$ & $6.2^{+2.0}_{-1.3}$\\
    $\alpha_\mathrm{out}$ & $-7.5^{+1.3}_{-2.2}$\\
    PA [$^{\circ}$] &  $119\pm4$ \\
    $i$ [$^{\circ}$] & $35\pm2$ \\
    %% RA offset [$\arcsec$] & $-0.04\pm0.09$  \\
    %% Dec offset [$\arcsec$] &  $0.01\pm0.07$ \\
    %% \\%& $0.482\pm6\times10^{-3}$\\ %&
    \hline
  \end{tabular}
\end{table}

%% Despite the fact that the model surface density displays
%% asymmetric features such as a spiral arms, we find that the simulated
%% observations are not sensitive enough to recover this structure, and
%% thus, they are roughly consistent with an axisymmetric disc. However,
%% even though the total flux is consistent with the one measured by
%% SCUBA-2/JCMT at 0.85 mm \citep[$15.5\pm1.4$ mJy,][]{Duchene2014}, the
%% LHB best fit model is significantly broader and hence has lower
%% surface brightness at 150 AU than the observed disc, leaving
%% significant positive and negative residals (see Figure
%% \ref{fig:res_profile}).

%% We vary the disc flux of to fit the ALMA observations. %% We find the
%% observations are
%% best fit with a total flux of x, but this is inconsistent with total
%% flux measured by SCUBA-2/JCMT at 0.85 mm \citep[$15.5\pm1.4$
%%   mJy][]{Duchene2014}.

%% 15.5 mJy measured by
%% SCUBA-2/JCMT \citep{Duchene2014}. Figure \ref{fig:models} c, f and i
%% show the synthetic continuum image at 0.88 mm, ALMA simulation and
%% dirty map of the residuals after subtracting it to the observed
%% visibilities. Under these assumptions, the LHB model disc is
%% significantly broader and fainter than the observed disc. However,
%% despite the fact that the model surface density displays asymmetric
%% features such as a spiral arms, we find that the simulated
%% observations are not sensitive enough to recover this structure, and
%% thus, they are consistent with an axisymmetric disc.

%% the LHB the outer disc is expected to be asymmetric \citep{Pearce2014}
%% and broad \citep{Booth2009}.

\subsection{Eccentric belt}
\label{sec:eccentric}
We also study the possibility of the belt being eccentric. This has
two main effects: the center of the belt is offset from the stellar
position, and the azimuthal density profile changes as particles spend
more time at the apocentre increasing its density relative to the
pericentre. This is known as apocentre glow
\citep{Pan2016}. Therefore, we fit a modified belt model in which we
replace $\rho_0$ and $r_{0}$ in Eq. \ref{eq:gausring} by
\begin{eqnarray}
  \rho_c&=&\rho_0 [1-e\cos(f)], \\
  r_c&=&\frac{r_0(1-e^2)}{1+e\cos(f)},
\end{eqnarray}
where $e$ is the eccentricity of the belt and $f$ is the true anomaly,
i.e. the azimuthal angle in the plane of the disc measured from the
direction of pericentre as seen from the stellar position.  We
maintain fixed the scale height to the best fit values presented above
and we added as free parameters $e$ and the position angle of the
pericenter, $w$, i.e. the angle in the plane of the sky between the
north direction and that of the pericentre as seen from the stellar
position, to have a total of 7 free parameters. We initiated the MCMC
using a uniform distribution of $w$ from $0^{\circ}$ to 360 $^{\circ}$
and a normal distribution for $e$ centered at 0 with a standard
deviation of 0.03. We run 10 MCMC routines in parallel, each one with
18 walkers to ensure a good sampling of the parameter space. We find
that the disc is consistent with being circular and from the posterior
distribution of $e$ we derive a 99.7\% upper limit of 0.05 (see
Sec. \ref{sec:displanet} for implications of this upper limit). We
find that the marginalised distribution of $w$ is not uniform and
constrained between $-80^{\circ}$ and $60^{\circ}$ (99.7\%
confidence), i.e. the pericentre if eccentric is more likely to be on
the northern half of the disc. However, $e$ still peaks at zero even
for these values of $w$.

%% which Therefore, we conclude that the difference on total flux of
%% the north-west and south-east halves, if real, is not produced by
%% apocentre glow due to the disc being eccentric.}

%% , where there is marginal evidence of being fainter, and thus,
%% consistent with apocentre glow.

\subsection{Model comparison}
\label{sec:modelcomparison}
\begin{figure}
  \includegraphics[trim=0.0cm 0.0cm 0.0cm 0.0cm, clip=true,
    width=1.0\columnwidth]{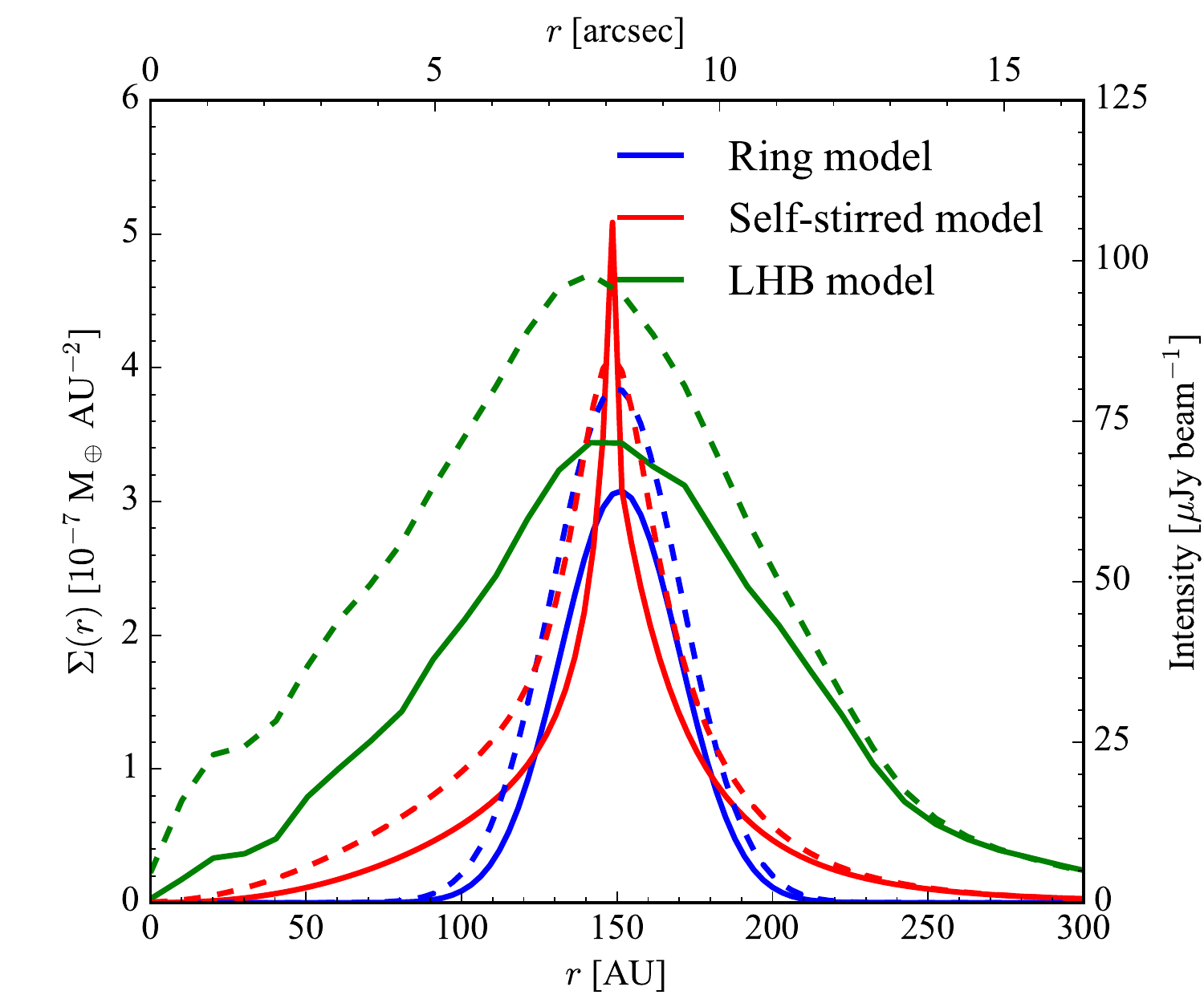}
  \caption{Dust surface density (continuous line) and surface
    brightness profiles convolved with a beam of $1\arcsec$ (dashed
    line) of the best fit belt, self-stirred and LHB models,
    represented in blue, red and green, respectively.}
    \label{fig:allmodels}
\end{figure}

\begin{figure}
  \includegraphics[trim=0.0cm 0.0cm 0.0cm 0.0cm, clip=true,
    width=1.0\columnwidth]{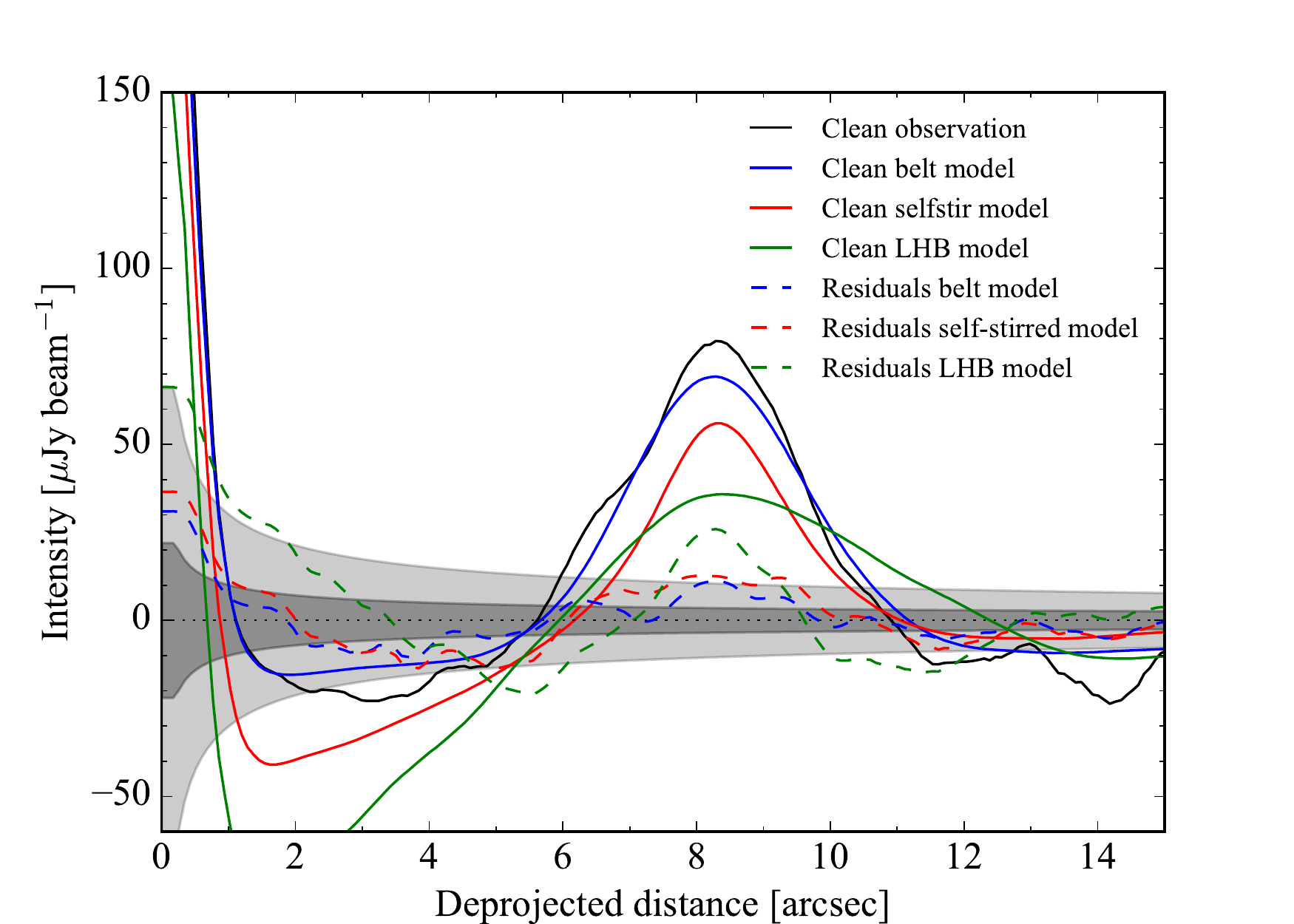}
  \caption{Azimuthally averaged intensity profiles obtained from CLEAN
    non tapered images of simulated observations without noise using
    the best fit models (continuous lines) and from the non tapered
    dirty map of the residuals after subtracting the best fit models
    (dashed). The blue, red and green lines represent the belt,
    self-stirred and LHB models, respectively. The \etacorvi \ CLEAN
    azimuthally averaged profile is represented with a continuous
    black line. The light and dark grey areas represent the 68\% and
    99.7\% confidence region.}
    \label{fig:res_profile}
\end{figure}

In Figure \ref{fig:allmodels} we compare the surface density and
surface brightness profiles of the belt and self-stirred model using
the best fit parameters defined above.  They display strong
differences in their surface density and brightness profiles, with the
self-stirred and LHB models extending significantly inside 100 AU and
outside 200 AU. Figure \ref{fig:res_profile} shows intensity radial
profiles obtained by averaging azimuthally simulated CLEAN images of
the best fit models without noise, and from the dirty maps of their
respective residuals when comparing with the observations. The three
models suffer from strong negative artefacts inside the cavity, with
the LHB model being the most extreme. Moreover, the self-stirred and
LHB models appear dimmer on the simulated observation than it should
based on the model (peak of 85 and 95 $\mu$Jy~beam$^{-1}$ at
$8\arcsec$ or 150 AU, respectively). This is probably due to an
insufficient number of short baselines to capture the broad
emission. In the same figure, we notice that the negative artefacts on
the CLEAN image are significantly reduced in the residuals below
$\pm3\sigma$ after subtracting the best belt model to the
visibilities, as expected if they are artefacts created by the image
reconstruction. %% Further discussion on this is presented in
%% Sec. \ref{dis:negatives}.

%% The CLEAN profiles of from the simulated observations
%% present differences between them that can be simply
%% attributed to the image noise. This is probably due to an insufficient
%% number of short baselines to quantify the more extended
%% emission. Moreover, similar to the intensity radial profile presented
%% in Figure \ref{fig:Ir}, negative artefacts of the same magnitude are
%% also present between $2\arcsec$ and $4\arcsec$ from the star in the
%% simulated CLEAN images. %% However, the emission from the outer belt
%% is significantly lower in the simulated CLEAN images of the best
%% fit models when compared with the observed intensity profile of
%% \etacorvi.

With the belt model we obtain the best $\chi^2$, followed by the
double power law model. In addition, to compare the goodness of fit of
the three best fit models, it is useful to compute the Bayesian
information criterion \citep[BIC,][]{Schwarz1978} for each one, which
penalises for the number of free parameters. It is defined as
$\chi^2+k\log(N)$, where $k$ is the number of free parameters (6 for
the belt model, 8 for the self-stirred model and 7 for the double
power law model) and $N$ is the number of data points or
visibilities. A model is more preferred when its BIC is the lower,
with a difference larger than 10 considered to be strong evidence
\citep{Kass1995}. We find that the belt model's BIC is lower by
$\sim40$ compared to the self-stirred best model's BIC, and lower by
20 compared to the double power law model. Thus, we conclude the
simple Gaussian belt model gives a better fit to the observations,
even with a lower number of free parameters.
%% \textcolor{red}{there is not enough information in the data to justify
%%   a more complex surface density model with more parameters than the
%%   three ($M_d$, $r_0$ and $\Delta r$) needed for the belt model.}

\subsection{Missed extended emission inside the belt}
\label{sec:missedfluxin}
It is possible that there could be emission in the cavity missed by
ALMA due to a low surface brightness or even due to an insufficient
number of short baselines to recover more extended emission. Assuming
a constant dust opacity with radius, we can estimate how much dust
mass could be hidden at an undetectable level between the star and the
belt at 150 AU. The maximum surface density ($3\sigma$ upper limit)
can be defined as
\begin{equation}
  \Sigma(r)=\frac{\delta I(r)}{\kappa_\mathrm{abs} B_{\nu}(T(r))}
\end{equation}
where $\kappa_\mathrm{abs}=3.8$~cm$^{2}$~g$^{-1}$ is the absorption
opacity of grains in our models and $\delta I(r)$ is the 3$\sigma$
uncertainty of the azimuthally averaged intensity profile, computed
considering the noise in the image space, as well as the apodization
by the primary beam. We also assumed $T(r)=42 (r/150
\ \mathrm{AU})^{-0.5}$~K, derived from the dust temperature in our
models described above (assuming a uniform grain size distribution and
composition in the disc). Integrating $\delta I(r)$ or $\Sigma(r)$ we
find a total missing flux $\lesssim3.4$ mJy or a possible hidden dust
mass $\lesssim2.4\times10^{-3}$ M$_\oplus$ in the cavity ($3\sigma$
limits). %% This leads to a total flux of $7.7\pm1.3$ mJy for the belt
%% model, where the uncertainty has increased because it now considers the
%% possible emission below the detection levels inside the cavity. This
%% is marginally consistent with the SCUBA-2/JCMT extrapolated flux
%% assuming a spectral index of 3.0 ($3.1\sigma$ difference considering
%% absolute flux calibration uncertainties).

\begin{figure}
  \includegraphics[trim=0.5cm 0.5cm 0.0cm 0.0cm, clip=true,
    width=1.0\columnwidth]{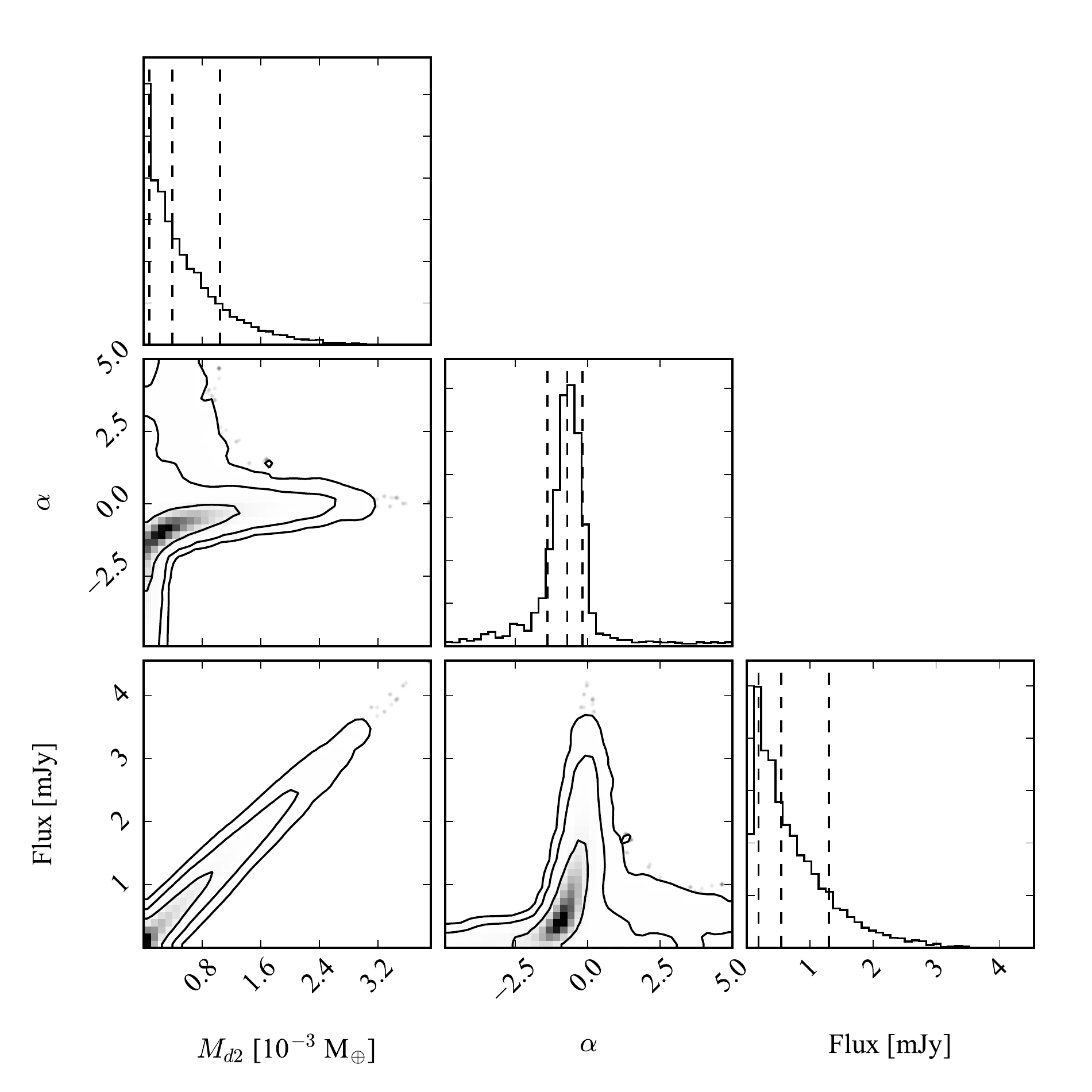}
  \caption{Posterior distribution of $\alpha$, $M_{d2}$ and the total
    disc flux of the second inner component. The vertical dashed lines
    represent the 16th, 50th and 84th percentiles. Contours correspond
    to 68\%, 95\% and 99.7\% confidence regions. This plot was
    generated using the python module \textit{corner}
    \citep{cornerplot}. }
    \label{fig:seccomp}
\end{figure}

However, this approach does not take into account the uv-filtering
that could be present due to missing short baselines, as the maximum
recoverable scale is $\sim7\arcsec$. Our observations are blind to
structure of that size or bigger. To study this effect and get a more
accurate constraint on the possible hidden emission inside the cavity
we added a second component to the simple belt model described in
Sec. \ref{sec:modelring} and we fit this to the observed
visibilities. The second component is defined with a surface density
proportional to $r^{\alpha}$ between 1 AU and 150 AU (inside the belt
mean radius), where we leave $\alpha$ as a free parameter that can
vary between -5 and 5, together with the total dust mass of the second
component, $M_\mathrm{d2}$.  In Figure \ref{fig:seccomp} we present
the distribution of $\alpha$, the total dust mass M$_\mathrm{d2}$ and
flux of the second component.  We find an $3\sigma$ upper limit of
$2.7\times10^{-3}$ M$_\oplus$ for $M_{d2}$ and 3.2 mJy for the flux of
the second component. This is similar to our previous analytic
estimate, which rejects the hypothesis of extended disc emission in
the cavity at detectable levels, but filtered out by missing short
baselines.  It is worth noting that this parametric method to
constrain the flux from the cavity assumes a power law surface density
distribution.

%% Moreover, from previous SCUBA-2/JCMT
%% observations at 0.85 mm \citep{Duchene2014} we know that there is
%% missing flux in our ALMA observations, probably due to missing short
%% baselines. From the fit of the flat component, we can also estimate a
%% $3\sigma$ upper limit of 12.1 mJy for the total disc flux at 0.88 mm,
%% marginally consistent with the SCUBA-2/JCMT flux ($\sim4\sigma$
%% difference).

\subsection{CO mass constraints}
\label{sec:modelco}
As shown by \cite{Matra2015} local thermodynamic equilibrium (LTE)
does not necessarily apply to the low density environments of debris
discs. Thus, when deriving gas masses from observed CO emission or
even upper limits from non detections, it is necessary to consider
non-LTE effects when the densities of the main collisional partners,
which we assume to be electrons, are not high enough. This choice is
justified because if CO gas of secondary origin is present, electrons
should be released from carbon ionization after the photodissociation
of CO, dominating the collisional excitations and de-excitations of
the CO rotational levels. However, we stress that the mass limits
derived in the radiation dominated regime (low electron densities) and
in LTE (high electron densities) are independent of the specific
collisional partner. Similar to Fomalhaut \citep{Matra2015}, the
radiation at 345 GHz in the disc is dominated by the CMB photons
($J_\nu=1.4\times10^{8}$ Jy~sr$^{-1}$) as the mean dust intensity
inside the belt is $\lesssim8\times10^{7}$~Jy~sr$^{-1}$ (calculated
from our best fit model in Sec. \ref{sec:modelring}) and the stellar
flux is $\sim10^{8}$~Jy~sr$^{-1}$ at 5 AU, which decreases steeply as
$r^{-2}$ and becomes negligible at 20 AU. We also neglect the flux
from the hot dust, as it is highly unconstrained at this
wavelength. We use the code developed by \cite{Matra2015} that
computes the population of the rotational levels in the non-LTE regime
to derive CO total fluxes and gas masses for different electron
densities (\nel) and gas kinetic temperatures ($T_\mathrm{k}$), assuming the
emission is optically thin.

\begin{figure}
  \includegraphics[trim=0.0cm 0.0cm 0.0cm 0.0cm, clip=true,
    width=1.0\columnwidth]{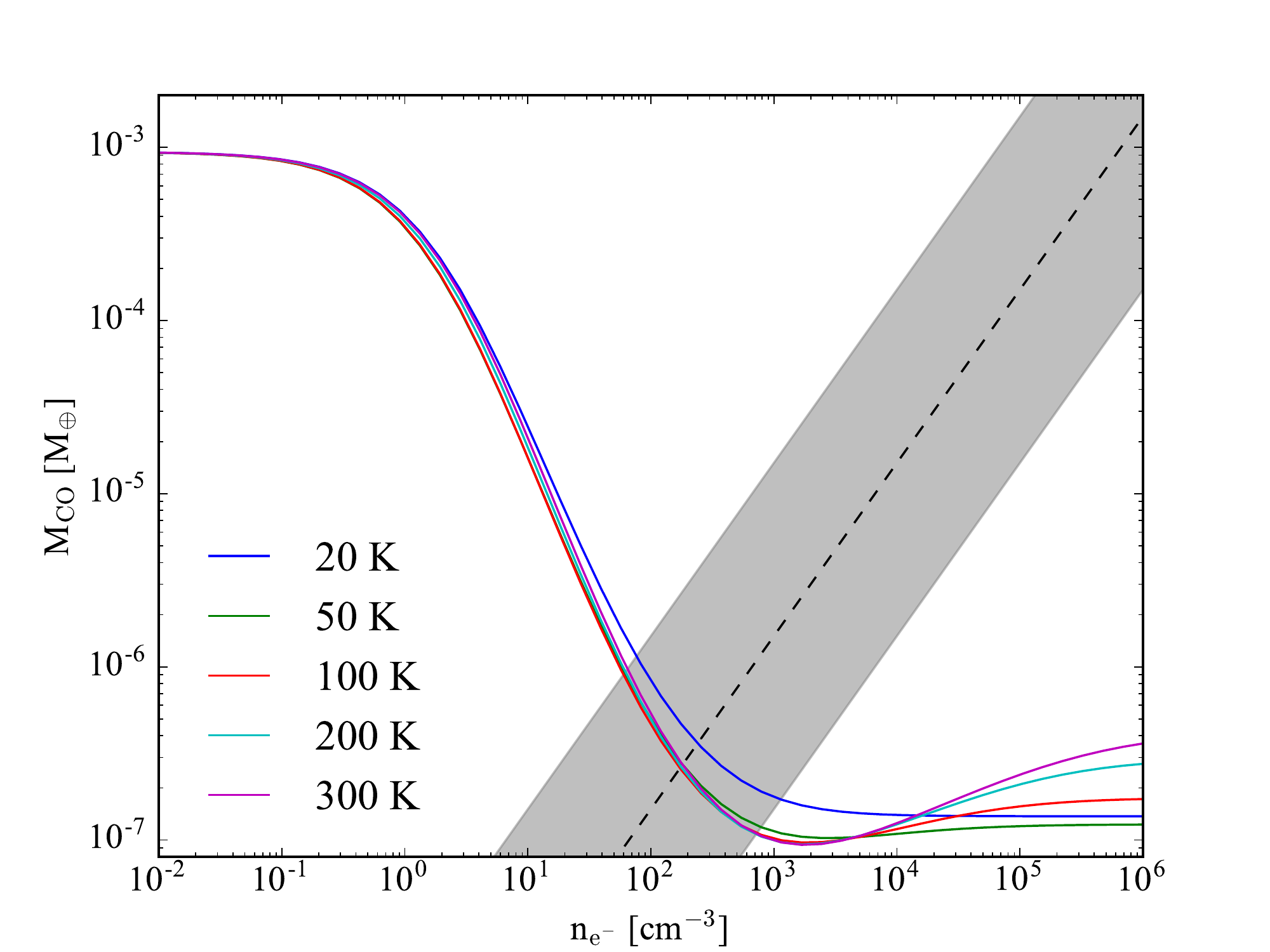}
  \caption{CO gas mass estimate of the detection at $\sim20$~AU for
    different gas kinetic temperatures and electron densities
    (n$_\mathrm{e^{-}}$ , main collisional partner). The dashed line
    corresponds to the CO gas mass assuming an abundance ratio of
    neutral carbon and CO of 100 and a carbon ionization fraction of
    0.5. The shaded grey region represents a factor of 10 of
    uncertainty on the CO and electron abundance ratio.}
    \label{fig:Mco}
\end{figure}

From the total flux of $38\pm9$~mJy~\kms measured between 15 AU and 37
AU (see Sec. \ref{sec:co}), we can estimate the CO gas mass located
around 20 AU considering different n$_\mathrm{e^{-}}$ and
$T_\mathrm{k}$ (20-300 K). This is shown in Figure \ref{fig:Mco}. As
expected from the work by \cite{Matra2015}, the mass of CO is highly
unconstrained without any knowledge of
n$_\mathrm{e^{-}}$. $M_\mathrm{CO}$ could easily be between
$\sim10^{-3}$ M$_\oplus$, in the radiation dominated regime (low
\nel), and $10^{-7}$ M$_\oplus$, in LTE (high \nel). On the other
hand, different gas kinetic temperatures can change $M_\mathrm{CO}$ by
a factor of a few when close to LTE.

However, in the secondary origin scenario electrons are a byproduct of
CO photodissociation and carbon ionization, therefore
n$_\mathrm{e^{-}}$ can be linked to M$_\mathrm{CO}$. In fact, Eq. 14
in \cite{Matra2015} shows that
\begin{equation}
  M_\mathrm{CO}=\frac{n_\mathrm{C}}{n_\mathrm{C^{+}}}
  \frac{n_\mathrm{CO}}{n_\mathrm{C}} \ m_\mathrm{CO} \ V_\mathrm{disc} \ n_\mathrm{e^{-}}\approx 0.01 \ m_\mathrm{CO} \ V_\mathrm{disc} \ n_\mathrm{e^{-}}, \label{eq:mcone}
\end{equation}
where m$_\mathrm{CO}$ is the mass of the CO molecule and
$V_\mathrm{disc}$ is the volume of the disc from 15 to 37 AU assuming
a vertical aspect ratio of 0.1. Based on previous studies of the
$\beta$ Pictoris disc, we assume $n_\mathrm{C}/n_\mathrm{C^{+}}=1$
\citep{Cataldi2014} and $n_\mathrm{CO}/n_\mathrm{C}=1/100$
\citep{Roberge2000} as first approximation. Therefore, if the CO gas
is of secondary origin, its mass is defined roughly by the
intersection of the dashed and continuous lines in Figure
\ref{fig:Mco}, i.e. $M_\mathrm{CO}\sim3\times10^{-7}$ M$_\oplus$. This
could vary by a factor of a few given the flux uncertainty and the
assumptions on $V_\mathrm{disc}$, the carbon ionization and CO/C
abundance ratio. These uncertainties are represented by the grey
shaded region defined between $0.1-10\ M_\mathrm{CO}$. We also find a
CO mass loss rate of $\sim3\times10^{-9}$ M$_\oplus$~yr$^{-1}$, as CO
photodissociates in a timescale of 120~yr due to the interstellar UV
radiation field \citep{Visser2009}, which is of the same order as the
hot dust mass loss rate \citep[$\sim3\times10^{-9}$
  M$_\oplus$~yr$^{-1}$,][]{Wyatt2010}. Possible origins of the CO gas
are discussed in Sec. \ref{dis:co}.

\begin{figure}
  \includegraphics[trim=0.0cm 0.0cm 0.0cm 0.0cm, clip=true,
    width=1.0\columnwidth]{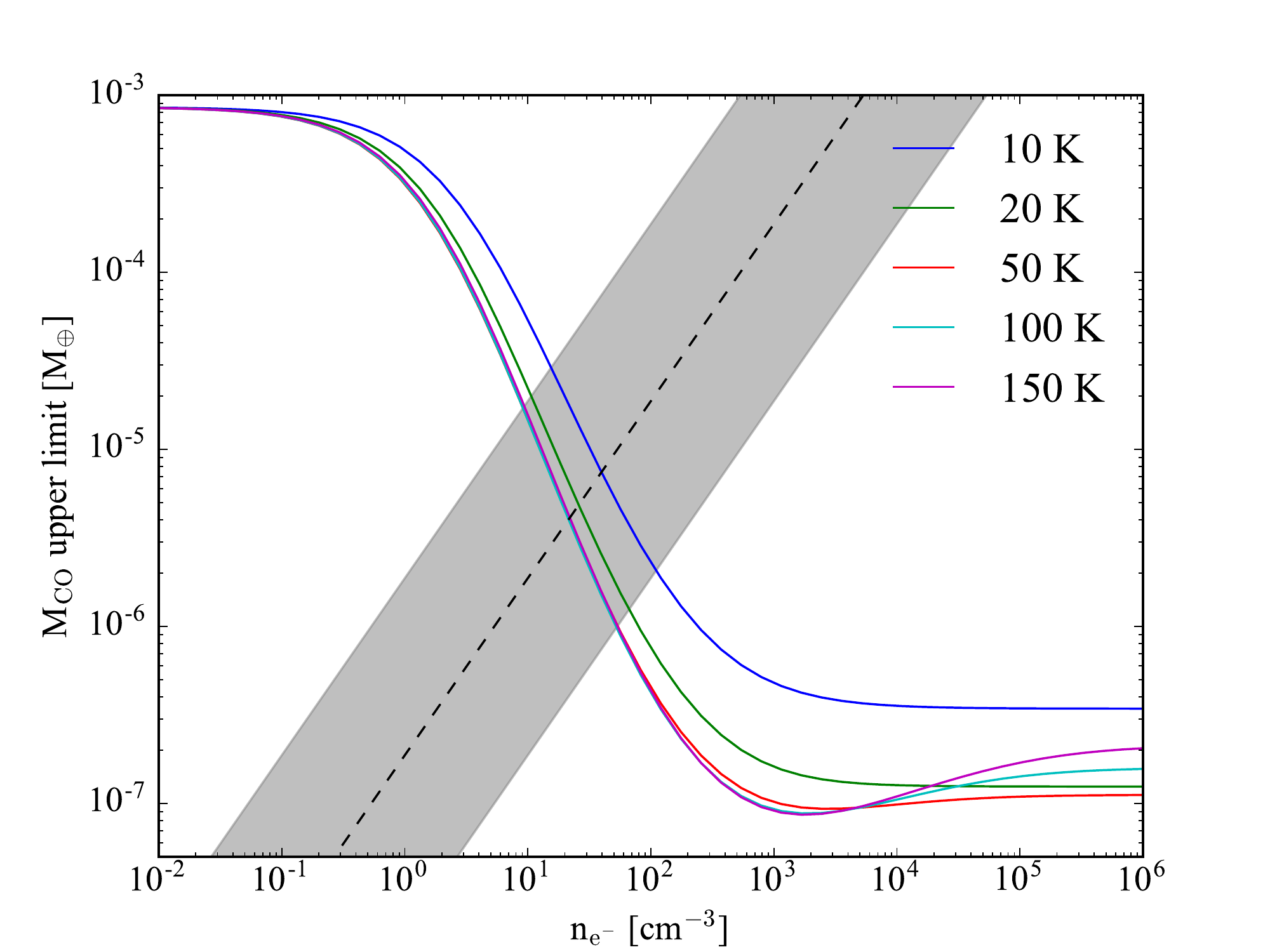}
  \caption{CO gas mass upper limits in the outer belt for different
    gas kinetic temperatures and electron densities
    (n$_\mathrm{e^{-}}$ , main collisional partner). The dashed line
    corresponds to the mass upper limit of CO gas assuming an
    abundance ratio of neutral carbon and CO of 100 and a carbon
    ionization fraction of 0.5. The shaded grey region represents a
    factor of 10 of uncertainty on the CO and electron abundance
    ratio.}
    \label{fig:Mcooutbelt}
\end{figure}

Based on the flux upper limits derived in Sec. \ref{sec:co}, we can
also estimate a mass upper limit on the CO gas mass that could be in
the outer belt or co-located with the hot dust. In Figure
\ref{fig:Mcooutbelt} we present the $3\sigma$ mass upper limits for
the CO in the outer belt, considering different $T_\mathrm{k}$ (10-150
K, a more appropiate range for $T_\mathrm{k}$ at 150 AU) and electron
densities. Similar to the derived mass above, the CO gas mass upper
limit is not well constrained and strongly depends on the density of
electrons in the disc varying by 4 orders of magnitude from $10^{-7}$
to $10^{-3}$ M$_{\oplus}$. Using Eq. \ref{eq:mcone} and the volume of
our best fit belt model we can estimate the expected density of
electrons for a given CO gas mass assuming the same carbon ionization
and C/CO abundance ratio observed in $\beta$~Pic. This is represented
with a dashed line. As the C/CO abundance and carbon ionization ratio
could be different than in $\beta$~Pic, we represent these
uncertainties with a the grey shaded region defined between
$0.1-10\ M_\mathrm{CO}$. The intersection between the dashed line and
the continuous lines in Figure \ref{fig:Mcooutbelt} gives the best
mass upper limit in the secondary origin scenario, which is
$4\times10^{-6}$ M$_{\oplus}$ for $T_\mathrm{k}=50$~K. The CO mass
could be higher than this limit, but this requires a low abundance of
$e^{-}$, which would be unlikely given the derived mass of CO, which
would photodissociate producing carbon that would get ionised
releasing further $e^{-}$ \citep{Kral2016}.

\begin{figure}
  \includegraphics[trim=0.0cm 0.0cm 0.0cm 0.0cm, clip=true,
    width=1.0\columnwidth]{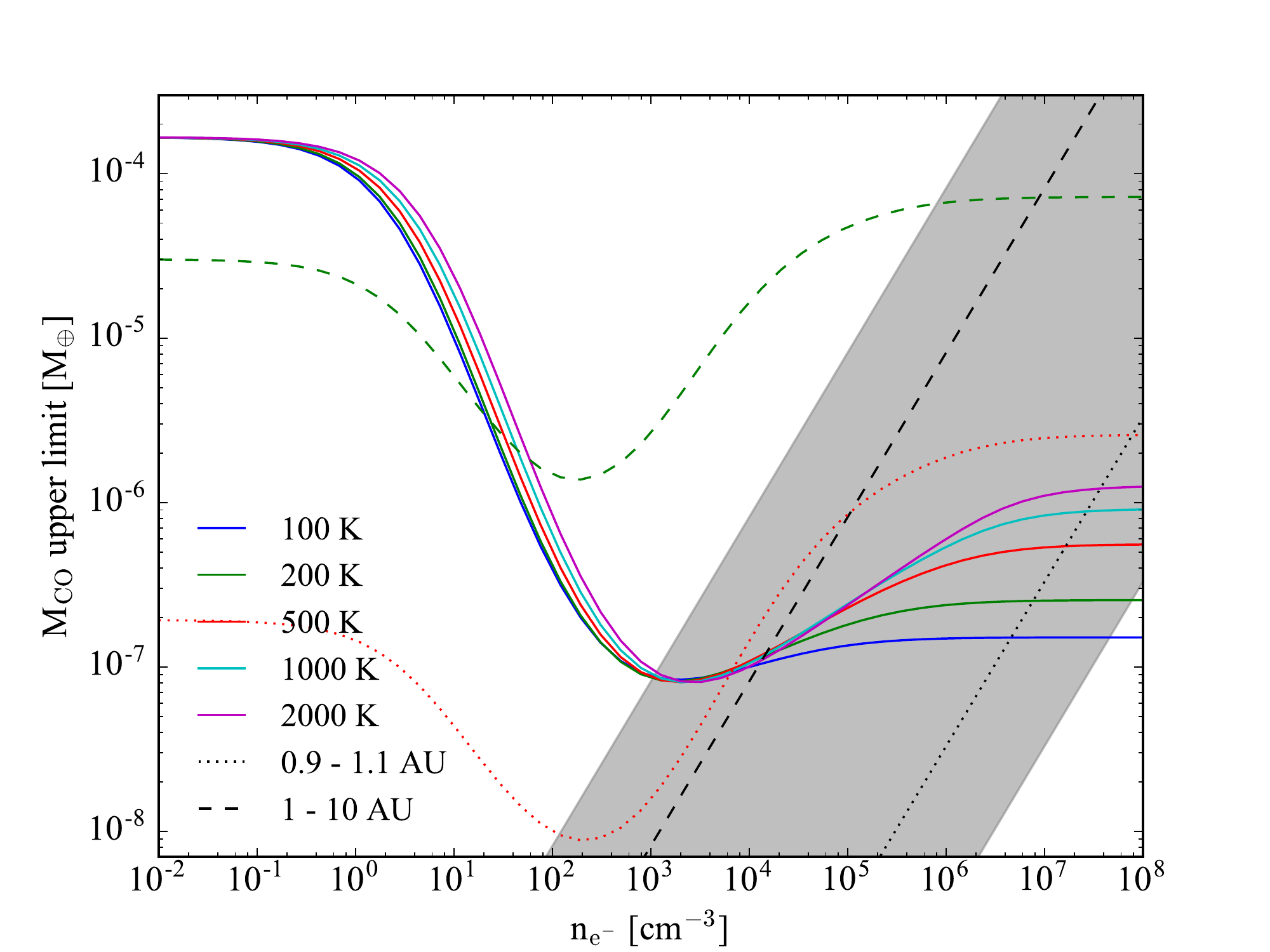}
  \caption{Mass upper limits of CO gas co-located with the hot dust
    for different gas kinetic temperatures and electron densities
    (n$_\mathrm{e^{-}}$, main collisional partner). The dashed and
    dotted black lines correspond to the mass upper limit of CO gas
    assuming a narrow ring and a broad inner disc, respectively, and
    an abundance ratio of neutral carbon and CO of 100 and a carbon
    ionization fraction of 0.5. The dashed and dotted curves represent
    the mass of CO gas at which the line (3-2) becomes optically thick
    ($\tau=1$) for a thin ring with $T_K=500$ K (red) and a broad
    inner disc with $T_K=200$ K (green). The shaded grey region
    represents a factor of 10 of uncertainty on the CO and electron
    abundance ratio, considering a range of volumes from a narrow to a
    broad inner disc.}
    \label{fig:Mcohotdust}
\end{figure}

Similarly, we calculate mass upper limits for the CO co-located with
the hot dust using the $3\sigma$ flux upper limit ($3\times11$
mJy~\kms), considering different $T_\mathrm{k}$ (100-2000 K, a more
appropiate range for $T_\mathrm{k}$ at a few AU) and including the
stellar flux at 345 GHz and at a radius of 1 AU (important for the
radiation dominated regime). This is shown in Figure
\ref{fig:Mcohotdust}. We also overlay in dashed and dotted black lines
the mass of CO gas as a function of the electron density assuming the
same carbon ionization fraction and C/CO abundance ratio as in
$\beta$~Pic. The dashed line corresponds to a disc with uniform
surface density that extends from 1 AU to 10 AU in radius, while the
dotted line represents a scenario in which most of the emission comes
from a narrow ring between 0.9 to 1.1 AU. Moreover, as the C/CO
abundance and carbon ionization ratio could be different than in
$\beta$~Pic, we represent these uncertainties with a grey shaded
region defined between $0.1
M_\mathrm{CO}^\mathrm{narrow}-10\ M_\mathrm{CO}^\mathrm{broad}$. The
intersections between the dashed and dotted lines with the continuous
line give the best estimate of the CO mass upper limit assuming a
secondary origin as in $\beta$~Pic. These two cases can be thought as
extremes cases, with the narrow ring scenario having the most
conservative upper limit of $\sim5\times10^{-7}$ M$_\oplus$ for
$T_\mathrm{k}=500$~K, roughly the equilibrium temperature at 1
AU. This limit is of the order of the CO gas mass derived before
located at $\sim20$ AU. However, more CO gas could be hidden if it
were distributed in an optically thick narrow ring. In the same figure
we added the mass of CO at which the CO line becomes optically thick
(mean $\tau_\nu$=1 across the line width), which is represented in red
dotted and green dashed lines for the two scenarios detailed above. We
find that the upper limits are optically thin in the regime we are
interested (secondary origin), unless the CO gas electron density is
very low and the CO in the disc is far from LTE.

If CO is released in collisions of icy planetesimals, e.g. releasing
CO gas trapped inside them or by exposing icy surfaces that can
sublimate via thermal or photodesorption, then the rate at which CO
gas is produced in the disc is given by
\begin{equation}
  \dot{M}^{+}_\mathrm{CO}=f_\mathrm{CO} \dot{M},
\end{equation}
where $f_\mathrm{CO}$ is the mass ratio of CO in planetesimals and
$\dot{M}$ is the mass loss rate of planetesimals. In steady state,
$\dot{M}^{+}_\mathrm{CO}$ is equal to the rate at which CO is lost in
the disc, i.e. $M_\mathrm{CO}/\tau_\mathrm{CO}$, where
$\tau_\mathrm{CO}$ is the photodissociation timescale given by the
interstellar UV radiation field \citep[$\sim120$~yr,
][]{Visser2009}. Using Eqs. 15 and 16 from \cite{Wyatt2008}, the
fractional luminosity of the outer disc
\citep[$\sim2\times10^{-5}$][]{Duchene2014}, its mean radius and width
derived in Sec. \ref{sec:modelring}, and assuming mean planetesimal
eccentricities of 0.05 and a uniform planetesimal disruption threshold
($\Qd$) of 200~J~kg$^{-1}$, we can estimate the outer disc mass loss
rate and what would be the CO gas mass in the outer belt for CO mass
fractions of planetesimal of 16\% \citep[maximum fraction derived in
  solar system comets,][]{Mumma2011}. We find that for these
parameters, $\dotM\sim10^{-3}$~M$_\oplus$~Myr$^{-1}$ (or
$\sim4\times10^{-5}$~M$_\oplus$~Myr$^{-1}$ for
$\Qd\sim10^{4}$~J~kg$^{-1}$) and
M$_\mathrm{CO}\sim3\times10^{-8}$~M$_\oplus$ or much lower if we
consider higher disruption thresholds or lower abundances of CO in
planetesimals. This mass of CO gas is much lower than our upper limit
of $5\times10^{-6}$ M$_{\oplus}$ derived above. The non-detection in
the outer belt is therefore consistent with Solar System comet
compositions.

On the other hand, the inner disc possess a fractional luminosity of
$\sim3\times10^{-4}$ \citep{Duchene2014}. Under the same assumptions
detailed above and including the stellar radiation, we find that for a
narrow ring $\dotM\sim2$~M$_\oplus$~Myr$^{-1}$ and there should be
$2\times10^{-5}$ M$_{\oplus}$ of CO gas, far above the upper limit
derived above for the inner disc. In fact, such a massive CO ring
would be optically thick ($\tau_\nu\sim7$). If we consider a broad
inner disc spanning from 1 to 10 AU, this limit decreases to
$2\times10^{-6}$ M$_{\oplus}$ ($\dotM\sim 0.1$~M$_\oplus$~Myr$^{-1}$),
but still above the upper limit on the broad inner disc
scenario. However, this prediction can be pushed down by increasing
the disruption threshold or decreasing the abundance of CO trapped in
planetesimals below 4\% \citep[still consistent with Solar System
  comets, $f_\mathrm{CO}$=0.3-16\%,][]{Mumma2011}. In fact, a low
fraction of CO gas trapped in planetesimals is expected if volatiles
have been lost at $\sim$20 AU on their way in from the outer belt (see
Sec. \ref{dis:co}).

%% \textbf{Moreover, the photodissociation
%% timescale at $r<1$ AU should be significantly shorter than 120 yr
%% as the stellar radiation starts to dominate the UV radiation
%% field \citep[computed from the photodissociation channels of CO,
%% ]{vanDishoeck1988}, and thus the predicted CO gas mass could be
%% lower by more than an order of magnitude.}

\section{Discussion}
\label{sec:dis}

\subsection{Hidden planet(s)}

\subsubsection{Belt or self-stirred disc}
\label{dis:selfstir}

In Sec. \ref{sec:model} we found that the disc continuum emission is
consistent with a belt at 150 AU and given the visibility
uncertainties and insufficient short baselines in our ALMA data, the
Gaussian belt model with three free parameters to describe the surface
density, gives the best fit and it is not necessary to invoke a more
complex model such as the self-stirred disc with two more free
parameters. Moreover, we found that the derived collisional timescale
of the biggest bodies in the self-stirred scenario, which controls the
width of the observed disc, is too short for a Pluto-sized
body. However, it could be that the primordial disc of planetesimals
was narrow. Then, it is no longer necessary that the stirring
timescale at 150 AU (i.e. the age of the system) is longer than the
collisional lifetime of the largest planetesimal, i.e.  $t_0\lesssim
t_\mathrm{age}$, to produce a narrow debris belt as the primordial
disc was already narrow; therefore, the outer belt could still be
self-stirred. A narrow distribution of planetesimals in the outer
belt, however it is stirred, could be due to a local enhancement of
solids at $\sim150$ AU caused by the presence of a snow line of a
specific volatile species that enhanced dust growth, or due to the
presence of a nearby planet truncating the inner edge of the outer
disc. In the latter scenario the planet could have stirred the disc
before self-stirring takes place \citep{Mustill2012}. It could also be
the case that the outer belt formed narrow without being truncated by
a nearby planet, but that a massive planet closer in stirred the outer
belt before self-stirring could take place.

Alternatively, the morphology could be due to radial dust trapping of
solids in pressure maxima at the edge of a gap or cavity during the
protoplanetary disc phase (i.e. when this would have been classified
as a transitional disc), where planetesimals could have grown. This
can be produced by the presence of a massive planet that opens a gap
in the disc \citep[e.g.,][]{Pinilla2012}, located around 100 AU in the
case of \etacorvi. For example HD~142527 has both an inner and outer
disc, with a cavity in both dust and gas extending from 10 AU out to
$\sim$140 AU \citep{Fukagawa2006, Casassus2012}. Although that system
has a low mass companion close in, it cannot have truncated the inner
edge of the outer disc \citep{Lacour2016}. Thus, a single or multiple
planets have been proposed to explain both the large cavity and the
transport of gas from the outer to the inner disc \citep[as suggested
  by ALMA observations,][]{Casassus2013Natur, Casassus2015twist},
necessary to maintain the accretion rate that otherwise would deplete
the inner disc in less than a year \citep{Verhoeff2011}. Similarly, to
explain the hot dust in \etacorvi \ a planetary system is required
passing in material from its outer belt. Such a planetary system would
probably stir the disc before self-stirring or the formation of a
Pluto-sized object takes place (see Sec. \ref{dis:max}).
  
%% Such a planet in \etacorvi \ would probably stir the disc before
%% self-stirring or the formation of a Pluto-sized object takes place,
%% and even scatter material from the outer belt to the inner regions as
%% proposed to explain its hot dust.

%% As this is the prefered scenario in \etacorvi \ to
%% explain the hot dust, we conclude that the disc is inconsistent with
%% self-stirring in a broad disc. This leaves open the possibility that
%% planets orbit interior to the belt, stirring the belt and transporting
%% material from it to within a few AUs.

%% being self-stirred.
%%  which is consistent with the hypothesis of a planetary system in
%% the disc, which could it could host planets stirring the disc that
%% could also transport material from the outer disc to within a few
%% AUs.

\subsubsection{From the outer to the inner disc}
\label{dis:outtoin}
% hot dust challenge and evindence

The two component debris disc around \etacorvi \ presents a challenge
to any theoretical predictions of debris discs as its hot dust cannot
be explained by a collisional cascade in situ as the system is too old
($\gtrsim1$ Gyr). Moreover, spectroscopic features of the hot dust
suggests that it was formed farther out, probably in the outer
belt. P-R drag alone is incapable of transporting enough dust from the
outer belt to the inner regions; therefore, a planetary system
scattering material from the outer belt is required. In
Sec. \ref{sec:intro} we identified three possible scenarios to explain
the hot dust in this system: 1) LHB-like instability, 2) a stable
planetary system scattering material and feeding a collisional cascade
closer in, and 3) a stable planetary system scattering planetesimals
that colliding with a planet within a few AU. These new ALMA
observations have shown that it is unlikely that the system is going
through an instability such as the LHB (scenario 1). Such a scenario
should leave asymmetric signatures, such as spiral arms or
stellar-disc offsets \citep[similar to the secular effect of an
  eccentric planet,][]{Pearce2014}, and a broad outer belt \citep[see
  Figure \ref{fig:lhb} from this work and Figure 1 in ][]{Booth2009}
during the evolution of the system. Given the rms achieved per beam in
these ALMA observations, we cannot discard asymmetric features of the
size of a few beams as they would not appear at a significant level
(compare Figures \ref{fig:models}c and \ref{fig:models}f). However, we
found that the outer belt is narrower compared with the LHB scenario
(See Sec. \ref{sec:modellhb}) or it has a much steeper surface density
slope within 150 AU, and that it has a small global eccentricity,
below 0.05. Thus, it seems unlikely that a dynamical instability in
the system similar to the LHB is responsible for the hot dust
excess. The outer belt has more likely retained a stable configuration
over Gyr timescales where icy material from the outer belt is being
passed in to the inner regions (scenarios 2 and 3).

\cite{Bonsor2012analytic} and \cite{Bonsor2012nbody} explored the
limits at which multiple planets on circular orbits within 150 AU can
scatter particles inwards from the outer belt. The second study showed
that it is difficult for planets more massive than Neptune to
transport high levels of material to the inner planetary system,
particularly after Gyr. This is based on the clearing of the planet's
chaotic zone of material. However, if this material is replenished,
planet scattering can sustain the hot dust. This can be done if the
putative planet is migrating outwards into the outer belt
\citep[e.g. by planetesimal scattering, ][]{Bonsor2014}.

Alternatively, solids could migrate inwards from the outer belt, e.g
via scattering with a low mass planet located in the middle of the
outer belt, through secular interactions with planets in the system,
or by P-R drag transporting small dust from the outer belt, that then
could fall into the chaotic zone of another planet located in the
cavity and possibly continue migrating inwards. The planet in the
middle of the belt should have a mass small enough such that the
clearing timescale of its chaotic zone is significantly longer than
the age of the system, i.e. $\lesssim 7 M_{\oplus}$ \citep[see Eq. 3
  in ][]{Shannon2016}, as the observed outer belt shows no evidence of
a gap or being cleared. Then, the icy material scattered from the
outer belt to the inner regions could feed the hot dust via mutual
destructive collisions of dust and planetesimals (scenario 2), or as a
product of giant impacts on a planet close which would release large
amounts of dusty debris (scenario 3). Although the exact rate at which
this material is transported and, thus, its radial distribution is
unknown, in Sec. \ref{sec:missedfluxin} we studied the possibility of
a shallow component in the disc with different surface density
distributions, connecting the outer belt with the hot dust region. We
constrained the surface density distribution as a function of its
slope and total dust mass. These limits can be used in the future to
test hypotheses of planet configurations that can deliver material
inward from the outer belt to the inner regions. For example, for a
surface density increasing with radius as $r^2$ in the cavity, it
cannot contain more dust mass than $8\times10^{-4}$~M$_\oplus$
(assuming $\kappa_\mathrm{abs}$=3.8 cm$^{2}$~g$^{-1}$) or a total flux
of 1 mJy.

Moreover, if the transport of material is in steady state, we can
place a lower limit on the rate at which solids are migrating inwards
by equating the hot dust mass loss rate with the inward flux of
solids. The first was estimated to be $\sim3\times
10^{-9}$~M$_\oplus$~yr$^{-1}$ \citep{Wyatt2010}. This implies that if
fed from the outer belt, this has lost at least $\sim4$~M$_\oplus$ in
the last 1.4 Gyr. On the other hand, the inward flux of solids can be
expressed as
\begin{equation}
  \dot{M}_\mathrm{hot}^{+}=2\ \pi\ r\ \Sigma(r)\ v_r(r),
\end{equation}
where $\Sigma(r)$ is the surface density of dust in the cavity and
$v_r$ is the migration rate from the outer belt to the hot dust
location. Here we assume $\Sigma(r)\propto r^{-1}$ and equal to
$10^{-8}(r/100\ \mathrm{AU})^{-1}$~M$_\oplus$~AU$^{-2}$, equivalent to
our upper limit of the total dust mass in the cavity
(10$^{-3}$~M$_\oplus$, see Sec. \ref{sec:missedfluxin}). Using the
lower limit on M$_\mathrm{hot}$ and upper limit on $\Sigma(r)$, we
find that $v_r\gtrsim5\times10^{-4}$~AU~yr$^{-1}$, i.e. a total
migration time of 0.25 Myr from 100 to 1 AU, assuming a surface
density distribution proportional to $r^{-1}$.  This lower limit (and
others for different $\Sigma(r)$) can be tested by N-body simulations
to assess scenarios 2 and 3 that could transport material from the
outer belt to the inner regions producing the hot dust.

\subsubsection{Constraints on a hidden plant at the disc inner edge}
\label{sec:displanet}

Although the depletion of dust interior to the outer belt and the high
hot dust excess in \etacorvi \ hint at the presence of stellar
companions or planets, searches for them have not been
successful. Radial velocity studies have discarded close in companions
down to the mass of a 6 Jupiter masses with a period shorter than 2000
days or with a semi-major axis less than 3 AU \citep{Lagrange2009,
  Borgniet2016}. Chandra and \textit{Spitzer} IRAC observations have
also discarded the presence of a sub-stellar companion that could
explain the unusually high X-ray luminosity of this old system
\citep{Marengoinprep}. However, given the new constraints on the disc
eccentricity, mean radius and width, we can put new constraints on a
hidden planet. Assuming there is a single planet inside the \etacorvi
\ cavity, it cannot have a large eccentricity, $e_\mathrm{plt}$, as an
eccentric planet would impose a forced eccentricity, $e_\mathrm{f}$,
on the disc through secular interactions. This forced eccentricity
must be lower than 0.05, the upper limit derived in
Sec. \ref{sec:modelring} for the disc eccentricity. The relation
between both eccentricities can be obtained from the disturbing
function. Here we adopt the expression with no restriction in
$e_\mathrm{plt}$, based on Eq. 8 in \cite{Mustill2009}
\begin{equation}
  e_\mathrm{plt}=
  -\frac{5\alpha}{8e_\mathrm{f}}+\sqrt{\left(\frac{5\alpha}{8e_\mathrm{f}}\right)^2+1}, \label{eq:ef}
\end{equation}
where $\alpha$ is equal to the ratio of the semi-major axis of the
planet, $a_\mathrm{plt}$, and the disc, $a_\mathrm{disc}=152$ AU.

\begin{figure*}
  \includegraphics[trim=0.0cm 0.0cm 2.0cm 0.0cm, clip=true,
    width=0.7\textwidth]{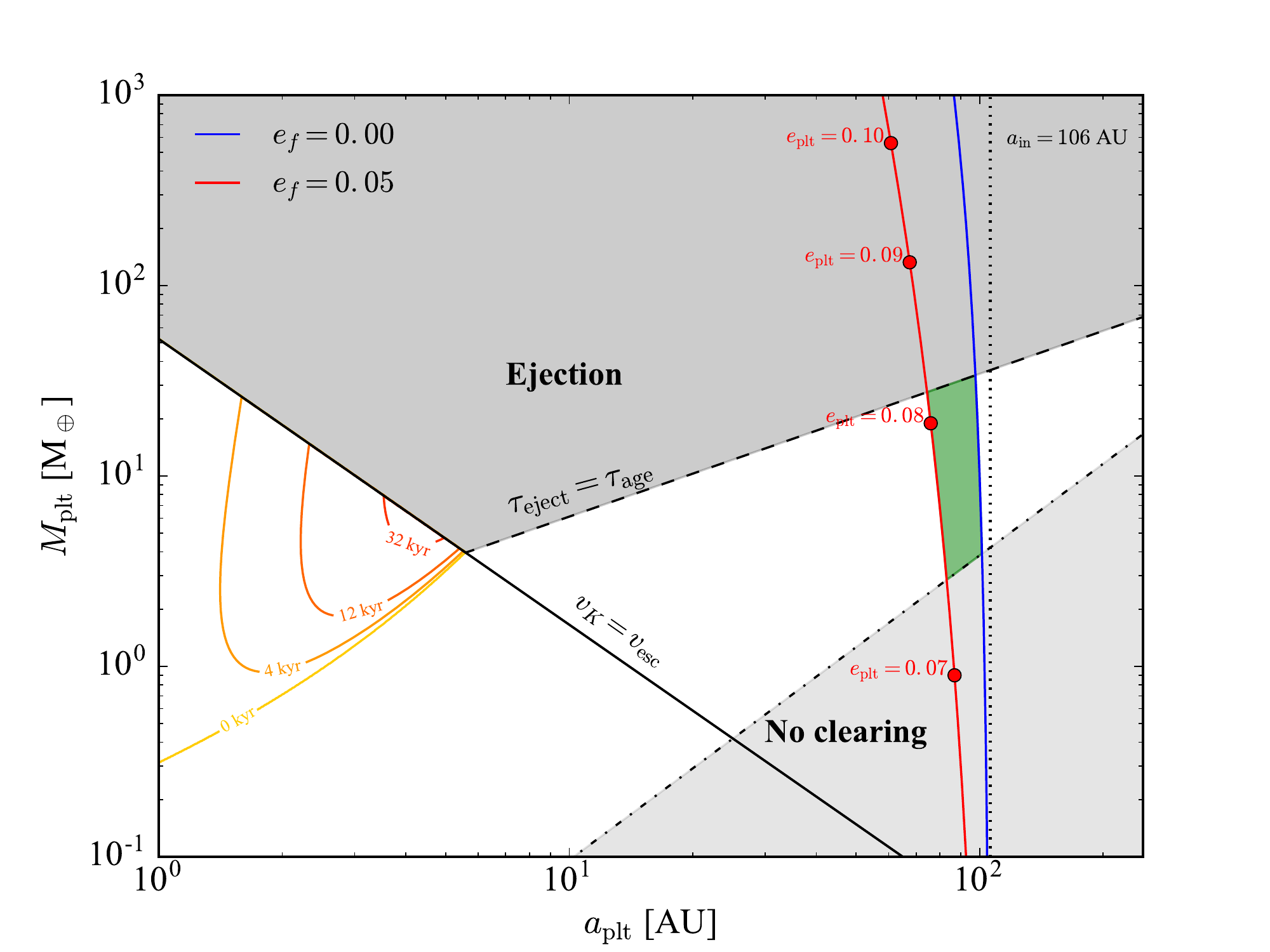}
  \caption{Constraints of the mass and semi-major axis of a perturbing
    planet in \etacorvi. The blue and red lines are defined by
    Eq. \ref{eq:mplt} using a forced eccentricity of 0 and 0.05,
    respectively. Eq. \ref{eq:meject} is represented by a continuous
    black line, while Eq. \ref{eq:mremain} is represented by a dashed
    black line. The mass of the planet to clear its chaotic zone on a
    timescale equal to the age of the system ($\sim1.4$ Gyr) is
    represented by a dashed dotted line. The red dots display the
    eccentricity of the planet that varies along the red curve,
    increasing with $M_\mathrm{plt}$. The inner edge of the disc is
    also shown with a dotted vertical line at 106 AU. The grey regions
    show excluded regions where the planet would be too massive
    ejecting most of the material it encounters or not massive enough
    to stir the disc. In green we highlight the region where
    $M_\mathrm{plt}$ and $a_\mathrm{plt}$ meet all the conditions
    above. The yellow, orange and red contours represent different
    combinations of $M_\mathrm{plt}$ and $a_\mathrm{plt}$ in which a
    giant impact would produce debris that can remain above a
    fractional excess of 0.5 at 20 $\mu$m for a certain timescale.}
  \label{fig:planet}
\end{figure*}

We further assume that this planet is also responsible for truncating
the outer disc and defining its inner edge, $a_\mathrm{in}$,
e.g. through direct scattering and overlap of mean motion resonance in
the so-called chaotic zone. The width of this zone has been estimated
analytically for both small \citep[Eq. 56 in ][]{Wisdom1980} and high
eccentricities \citep[Eq. 10 in ][]{Mustill2012}, which we can use to
relate the mass of the planet ($M_\mathrm{plt}$) and its semi-major
axis
%% \citep{Wisdom1980}, we can use Eqs. 6 and 10 from
%% \cite{Mustill2012} to relate the mass of the planet ($M_\mathrm{plt}$)
%% and its semi-major axis by
\begin{equation}
 M_\mathrm{plt}(a_\mathrm{plt},e_\mathrm{f}) = \begin{cases}
   \left(\frac{ \frac{a_\mathrm{in}}{a_\mathrm{plt}} -1 }{1.3}\right)^{7/2} \ M_\star \  & \text{$e_\mathrm{plt}<e_\mathrm{crit}$}
   \\
   \left(\frac{ \frac{ a_\mathrm{in}}{a_\mathrm{plt}} -1 }{1.8}\right)^{5} \ e_\mathrm{plt} \ M_\star \ & \text{$e_\mathrm{plt}>e_\mathrm{crit}$}
\end{cases} \label{eq:mplt}
\end{equation}
where $a_\mathrm{in}=r_0-\Delta r=106$ AU, $M_\star\sim1.4\ M_\odot$
and $e_\mathrm{crit}$ is the critical eccentricity at which
$e_\mathrm{plt}$ needs to be considered for the size of the chaotic
zone \citep[Eq. 11 in ][]{Mustill2012}, which for a Neptune mass
planet is $\sim0.002$. Eqs. \ref{eq:ef} and \ref{eq:mplt} define a
surface in $e_\mathrm{f}$ vs $a_\mathrm{plt}$ vs $M_\mathrm{plt}$ space where
this planet could reside. In Figure \ref{fig:planet} we present
$M_\mathrm{plt}$ as a function of $a_\mathrm{plt}$ for $e_\mathrm{f}=0$ and
0.05 in continuous blue and red lines, respectively. With red dots we
indicate the eccentricity of the putative planet along the
$M_\mathrm{plt}(a_\mathrm{plt})$ curve. Both lines are nearly vertical
and the region between them defines where a planet defining the outer
belt's inner edge could lie.

However, a low mass planet would clear its chaotic zone on timescales
of the order or even longer than the age of the system. We can use
Eq. 3 in \cite{Shannon2016} which defines the clearing timescale of
the chaotic zone as a function of the planet mass and semi-major axis,
to estimate the minimum mass of the planet sculpting the inner edge
\begin{equation}
  M_\mathrm{plt}(a_\mathrm{plt})=4 \left(\frac{t_\star}{1.4 \mathrm{Gyr}}\right)^{-1
  } \left(\frac{a_\mathrm{plt}}{100 \ \mathrm{AU}}\right)^{1.6} \left(\frac{M_\star}{1.4 \ \mathrm{M}_\odot}\right)^{-1/2} \ [\mathrm{M}_\oplus] \label{eq:mclear}
\end{equation}
A planet below this mass limit will not clear its chaotic zone fast
enough to truncate the inner edge of the outer belt. Therefore, the
planet sculpting the outer belt should lie between 60 and 110 AU and
have a mass higher than 3 M$_\oplus$.

%% \caption{Constraints of the mass and semi-major axis of a perturbing
%%     planet in \etacorvi. The blue and red lines are defined by
%%     Eq. \ref{eq:mplt} using a forced eccentricity of 0 and 0.05,
%%     respectively. Eq. \ref{eq:meject} is represented by a continuous
%%     black line, while Eq. \ref{eq:mremain} is represented by a dashed
%%     black line. The mass of the planet for which the stirring
%%     timescales is equal to the age of the system ($\sim1.4$ Gyr) is
%%     represented by a dashed dotted line. The red dots display the
%%     eccentricity of the planet that varies along the red curve,
%%     increasing with $M_\mathrm{plt}$. The inner edge of the disc is
%%     also shown with a dotted vertical line at 108 AU. The grey regions
%%     show excluded regions where the planet would be too massive
%%     ejecting most of the material it encounters or not massive enough
%%     to stir the disc. In green we highlight the region where
%%     $M_\mathrm{plt}$ and $a_\mathrm{plt}$ meet all the conditions
%%     above. The yellow, orange and red contours represent different
%%     combinations of $M_\mathrm{plt}$ and $a_\mathrm{plt}$ in which a
%%     giant impact would produce debris that can remain above a
%%     fractional excess of 0.5 at 20 $\mu$m for a certain timescale.}
%% \label{fig:planet}

\subsubsection{Maximising the inward flow}
\label{dis:max}
As commented before, if material from the outer belt can migrate
inwards, it could fall in the chaotic zone of the putative planet
sculpting the outer belt, possibly continuing its inward migration and
feeding the hot dust before being ejected or accreted. Assuming that
the hot dust is fed by planet scattering in this manner, we can place
constraints on the planet masses required, and thus, further constrain
the orbital parameters of any interior planet. As the maximum kick a
particle can experience when encountering a planet is of the order of
the planet's escape velocity, $v_\mathrm{esc}$, planets with
$v_\mathrm{esc}$ much larger than the Keplerian velocity,
$v_\mathrm{K}$, will most likely eject particles after multiple
kicks. On the other hand, if $v_\mathrm{esc}\ll v_\mathrm{K}$ then
accretion will likely be the final outcome before the particle gets
enough kicks to put it on an unbound orbit. As shown in
\cite{Wyatt2016scat}, equating $v_\mathrm{esc}$ and $v_\mathrm{K}$ we
can find roughly the planet mass that divides the two scenarios
%% devide the $M_\mathrm{plt}$-$a_\mathrm{plt}$ space the most natural
%% outcome of such particles can be particles encountering a planet
%% can get scattered, ejected or accreated. the most natural outcome a
%% planet of a given mass and semi-major axis would for a planet not
%% to eject the material it encounters it has too have a mass roughly
%% equal or lower than the planet mass defined equating the Keplerian
%% velocity with the escape velocity of such planet
\begin{equation}
  M_\mathrm{plt} = 40 \left(\frac{M_\star}{M_\odot}\right)^{3/2} \left(\frac{a_\mathrm{plt}}{1\ \mathrm{AU}}\right)^{-3/2} \left(\frac{\rho_\mathrm{plt}}{1\ \mathrm{g~cm^{-3}}}\right)^{-1/2} \ [M_\mathrm{\oplus}], \label{eq:meject}
\end{equation}
where $\rho_\mathrm{plt}$ is the bulk mass of the planet, hereafter
assumed to be 1.6~g~cm$^{-3}$ (Neptune's bulk density). This line is
represented in Figure \ref{fig:planet} by a continuous black
line. However, ejection might only happen after several encounters,
thus, material can remain in the system for timescales shorter than
the ejection timescale, which we take as the cometary diffusion
timescale derived empirically by \cite{Tremaine1993}. Using Eq. 2 from
\cite{Wyatt2016scat} we find
\begin{equation}
M_\mathrm{plt}=\left(\frac{M_\star}{M_\odot}\right)^{3/4} \left(\frac{a_\mathrm{plt}}{1\ \mathrm{AU}}\right)^{3/4} \left(\frac{t_\star}{1 \mathrm{Gyr}}\right)^{-1}  \ [M_\mathrm{\oplus}]. \label{eq:mremain}
\end{equation}
In Figure \ref{fig:planet} this is represented by a dashed black line
(using $t_\star$=1.4 Gyr). Therefore, particles encountering a planet
with a mass above the one defined by Eq. \ref{eq:meject} and below
Eq. \ref{eq:mremain} will likely remain in the system without being
ejected or accreted for timescales longer than the age of the
system. Thus, a planet that both truncated the disc (red and blue
lines) and does not eject particles within 1.4 Gyr must have a mass
$\lesssim30$ M$_\oplus$. However, material formed farther out, e.g. in
the outer belt, could fall in this region at late epochs and thus
remain in the system until the present epoch even if the putative
planet is more massive than 30 M$_\oplus$.

%% Finally, in Sec. \ref{sec:modelselfstirred} we found that the
%% self-stirred scenario is unlikely to be responsible for the stirring
%% in the \etacorvi \ outer disc; thus, the disc has to have been
%% stirred by another mechanism to produce the current levels of dust
%% through destructive collisions. Here, we propose that the disc has
%% been stirred by the presence of an unseen planet. This agrees with the
%% fact that the presence of planets is needed to explain how the outer
%% disc is feeding the inner regions ($\lesssim3$ AU) as explained
%% above. Planet stirring through secular perturbations has been studied
%% before. Typically the disc is considered stirred when orbits initially
%% circular starts to cross each other due to secular perturbations of an
%% eccentric planet. Assuming a forced eccentricity of 0.05, we use
%% Eq. 15 derived by \cite{Mustill2009} for small eccentricities to
%% define a lower limit of the planet stirring the disc
%% \begin{equation}
%%   \begin{split}
%%   M_\mathrm{plt}=0.93 \frac{(1-e_\mathrm{plt}^2)^{3/2}}{e_\mathrm{plt}}\left(\frac{a_\mathrm{disc}}{152\ \mathrm{AU}}\right)^{9/2}  \left(\frac{a_\mathrm{plt}}{100\ \mathrm{AU}}\right)^{-3} \left(\frac{M_\star}{1.5 M_\odot}\right)^{0.5} \\
%%   \left(\frac{\tau_\mathrm{age}}{1.4 \ \mathrm{Gyr}}\right)^{-1} \ [M_\mathrm{\oplus}].
%%   \end{split}
%% \end{equation}

Combining the four equations above we can define a region in
$M_\mathrm{plt}$ vs $a_\mathrm{plt}$ space where the putative planet
is most likely to be found. This is represented by a green region in
Figure \ref{fig:planet}, and roughly defined by $a_\mathrm{plt}$ in
the range $\sim$ 75-100 AU and $M_\mathrm{plt}$ between 3-30
$M_{\oplus}$. Planets in this region will have a semi-major axis such
that they can truncate the disc by clearing their chaotic zone on a
timescale shorter than the age of the system. They will not force an
eccentricity on the disc higher than 0.05, and they will not eject
particles within 1.4 Gyr. Given the uncertainty on the age of the
system (1-2 Gyr), the range of masses given above could change by
40\%. The mass upper limit is also consistent with the limits of a few
Jupiter masses placed by direct imaging planet searches in this
system.

Moreover, the planets in the green region have higher masses than the
minimum mass to stir the outer belt on a timescale equal to the age of
the system. Assuming a forced eccentricity of 0.05,
i.e. $e_\mathrm{plt}\lesssim0.08$, and using Eq. 15 from
\cite{Mustill2009} we find this mass is 1~$M_{\oplus}$. Therefore,
self-stirring is not necessary as a planet in this region would be
able to stir the disc on a timescale shorter than 1.4 Gyr.

\subsubsection{Multiple planets}

It is important to stress that our analysis above is only valid for a
single planet. The planet truncating the disc could have a higher mass
if other planets are present with smaller $a_\mathrm{plt}$, such that
particles initially scattered by the outermost planet get scattered
inward by the inner planets before being ejected. In a chain of
planets this can be achieved consecutively if each planet on the chain
has a scattering timescale shorter than the next planet with larger
semi-major axis, such that particles get scattered faster by the inner
planets, increasing the probability of migrating inwards. In other
words, the scattering timescale of the planets in the system has to
increase with distance in order to maximise the inward torque. This
could be done by a flat distribution of planet mass with
$a_\mathrm{plt}$ as the scattering timescale scales with orbital
period. However, a chain of planets with uniform mass above 30
M$_\oplus$ would scatter material inwards until reaching the innermost
planet which will eject particles in only a few encounters without
feeding the hot dust.
%% On the other hand, the outermost planet could have a mass smaller
%% than the minimum defined in the previous section as other planets
%% with a mass high enough to stir the disc could be present.

Therefore, as discussed in \cite{Wyatt2016scat} we can hypothesise how
a chain of planets should be distributed in mass and semi-major axis
to maximise the influx from the outer to the inner most regions. The
two basic requirements are: (1) the innermost planet within a few AU
should reside below the ejection region of Figure \ref{fig:planet},
such that particles can remain in the inner regions for longer
timescales or produce giant impacts releasing large amounts of dusty
debris as accretion is more likely; (2) the mass of planets should be
close to a flat distribution or decrease with radius, maximising the
inward torque. Based on Figure \ref{fig:planet} and the two conditions
above, we find that a chain of planets with uniform mass between
$3-30$~M$_\oplus$ would satisfy the two conditions above, e.g. a chain
of less than 10 planets separated by $\gtrsim$15 mutual Hill radii
with a mass of $\sim10$ M$_\oplus$.
%% planets down to 1 AU in a stable configuration, e.g. less than 10
%% planets separated by $\gtrsim$15 mutual Hill.
Such a system would scatter particles from the outer regions until
they encounter the innermost planet in the chain which would lie below
the ejection region. Then, the hot dust excess observed in \etacorvi
\ could be the product of giant impacts releasing debris or the
product of collisions between larger grains or planetesimals within a
few AU where collisions are more frequent and destructive.

%% \subsubsection{Planet-Planet scattering}
%% A third scenario that could explain the hot dust is if the system
%% contained a chain of several low mass planets in the cavity on an
%% unstable configuration over timescales of the order of the age of the
%% system \citep[see ][]{Smith2009}, e.g. twenty $\sim5$~M$_\oplus$
%% initially distributed between 1 and 100 AU. These planets initially
%% located above the ejection vs accretion line would get mutually
%% scattered until reaching the inner regions, at which accretion or
%% giant impacts are more likely. These giant collisions would release
%% large amount of debris that could explain the observed hot dust. In
%% this scenario no material needs to be transported from the outer belt
%% to the inner regions. \textbf{How does this fit with the CO detection?,
%%   can we say that the tentative detection could be from planets
%%   loosing part of their atmospheres in this scenario?}.

%% in which no material is being
%% transported from the outer disc to the inner regions, but unstable
%% over timescales of the order of the age of the system. These planets
%% would get mutually scattered until reaching the inner regions at which
%% accretion or giant impacts are more likely releasing large amount of
%% debris and producing the hot dust.

\subsubsection{The \etacorvi \ sweet spot}
\label{sec:sweetspot}

In the scenario 3, the hot dust in \etacorvi \ is the result of a
giant collision of an embryo scattered from the outer disc with a
planet at separations $\lesssim3$~AU. Based on the work by
\cite{Wyatt2016scat} we can place some constraint on this planet or
others within a few AU. In their work they estimated how long debris
produced from a giant collision can remain at detectable levels before
being accreted or collisionally depleted (see their Eq. 22 and Figure
5). In Figure \ref{fig:planet} we plot contours of the timescales at
which debris produced by a giant collision can remain above the
current levels in \etacorvi. We assumed that the giant impact puts in
orbit a fraction $f_\mathrm{esc}=0.05$ of the mass of the planet, that
evolves through mutual collision or it gets accreted by the
planet. The debris is characterised by a maximum particle size of 100
km and a planetesimal threshold $\Qd=200$~J~kg$^{-1}$. SED modelling
showed that the peak of the hot dust is around 20 $\mu$m with a
fractional excess ($R_{20}$) of about 0.5 \citep{Duchene2014}, thus we
are interested in the timescale at which the hot dust emission can
stay above $R_{20}$, $t_{>R_\lambda}$. These timescales are shown in
Figure \ref{fig:planet} with yellow, orange and red contours.

From Figure \ref{fig:planet} we find that if giant collisions are
uniformly likely in the log($a_\mathrm{plt}$)-log($M_\mathrm{plt}$)
space, an excess at 20 $\mu$m is more likely to be produced around 3-5
AU by a collision on a planet of 4-10~M$_\oplus$. This excess would
last around $10^4$~years or a few thousands orbits. Throughout this
time the debris would remain in an asymmetric distribution
\citep{Jackson2014}. Such asymmetry would naturally explain why the
emission is seen at a projected separation in the range 0.5-1 AU by
the LBTI \citep{Defrere2015}, but its temperature suggests a physical
separation of $\sim3$~AU where the collision would have
happened. Otherwise, this is hard to reconcile if the hot dust is
produced in an axisymmetric disc, where a collisional cascade is fed
from the outer belt (scenario 2). For scenario 2 to work this requires
a very high albedo and a grain size distribution steeper than the
expected from a collisional cascade \citep{Lebreton2016}, which thus
means that scenario 3 may be favoured.

\subsection{CO origin}
\label{dis:co}

Given the old age of \etacorvi \ ($\gtrsim1$ Gyr) any gas present in
the system has to be of secondary origin, i.e. released from icy
bodies present in the system. This could happen either in the outer
belt or in the inner disc if CO is released as a consequence of
collisions between icy bodies; however, the tentative detection of CO
is not co-located with either of the two. Recently \cite{Kral2016}
studied the evolution of secondary gas in a debris disc when produced
from a narrow ring. The gas can viscously spread forming an accretion
disc, with a surface density that depends on the gas production rate
and on the viscosity of the disc, expected to be higher than in a
protoplanetary disc as the disc should be highly ionised
\citep{Kral2016b}. Depending on the viscous timescale and lifetime,
different molecules or atomic species would have different
distributions in the disc. For example, in the case of $\beta$~Pic, CO
gas is released at $\sim85$ AU from icy bodies and photodissociates in
C+O, which then spread to form an atomic accretion disc. Atomic gas
species such as HI, CI, CII and OI, products of the photodissociation
of H$_2$O and CO can viscously spread in the disc before being
accreted into the host star \citep{Kral2016}. On the other hand, CO
has a photodissociation timescale of $\sim300$ yr \citep[longer than
  120 yr as it is slightly self-shielded,][]{Matra2016}, which is only
a fraction of an orbit, and then too short to be able to spread in the
disc. This is why CO is found to be co-located with the
millimetre-sized dust in the $\beta$~Pic disc, where it is released
\citep{Dent2014, Matra2016}.

However, if gas is released closer in, e.g. at $\sim1$ AU where the
hot dust is located in \etacorvi, then CO could spread significantly
outwards in the disc as both the orbital and viscous timescale
increase with radius. In order to assess if the CO could spread up to
20 AU, we can estimate the viscous timescale, $t_\nu$, for it assuming
an $\alpha$-parametrization for the viscosity \citep{Shakura1973},
i.e. $\nu=\alpha_\nu c_s \Hr$, where $c_s$ and $\Hr$ are the sound
speed and the local disc scale height. Assuming $\alpha_\nu=1.5$
\citep[best fit model for $\beta$~Pic,][]{Kral2016} we find that
$t_\nu\sim \tfrac{\Delta r^2}{\nu(r_c')}$ is $\sim10^4-10^5$ yr for
$\Delta r=20$ AU and $r_c'$=1-20 AU. This is longer by at least two
orders of magnitude than the photodissociation timescale, and thus, CO
would not be able to spread fast enough to reach 20 AU before being
photodissociated. If the CO production rate were high enough, CO could
be self-shielded; however, the upper limit we found in
Sec. \ref{sec:modelco} for a broad inner disc implies that the
vertical column density of CO is $\lesssim10^{14}$~cm$^{-2}$, not
enough to be considerably self-shielded \citep{Visser2009}. Note that
carbon ionization can also shield the CO, yet in $\beta$~Pic this was
found to be a minor effect compared to the CO self-shielding
\citep{Matra2016}. Assuming a carbon to CO abundance ratio of 100,
Carbon ionization would shield the CO only by a factor of 1.2,
increasing its photodissociation timescale to 140 yr
\citep{Rollins2012}. Another important issue with this scenario is
that within a few AU solids should be depleted of icy volatiles as the
temperatures are significantly higher than 140 K, the maximum
temperature at which CO or CO$_2$ can be trapped by amorphous H$_2$O
ice \citep{Collings2003}. In \etacorvi \ this temperature corresponds
to the equilibrium temperature at 9 AU. Therefore, icy material would
have to pass very quickly this ice line to reach $\sim$1 AU to release
the CO in destructive collisions of planetesimals or be large enough
as the sublimation rate depends on their area, e.g. comets in Solar
System that cross the water ice line can retain significant amounts of
ices for several orbits.

Alternatively, the tentative CO detection at $22\pm6$ AU could be
explained by gas released in situ. Icy material formed in the outer
belt and transported from cold regions, should start to sublimate at a
high rate when crossing a specific ice line or reaching specific
temperatures. At $22\pm6$ AU, the equilibrium temperature is $88\pm12$
K. Desorption experiments have studied in detail the desorption rate
of volatiles in isolation or with the presence of water ice, as a
function of the temperature. For example, in isolation the CO$_2$
desorption rate peaks at around 80 K \citep{Collings2004}, thus CO gas
trapped in CO$_2$ ice could be released when crossing the CO$_2$ ice
line; alternatively, as CO$_2$ photodissociates in about 30 yr into CO
and O due to the interstellar UV radiation field \citep{Hudson1971,
  Lewis1983}, the observed CO could all come from CO$_2$ being
photodissociated. For example, the coma activity of comet C/2012 S1
(ISON) on its last passage was likely controlled by CO$_2$ ice
sublimation beyond the water ice line \citep{Meech2013}. Another
possibility is that if the volatile content in planetesimals is
dominated by water ice, the desorption rates of other molecules can be
modified as they get trapped by amorphous water ice. This is true for
trapped CO and CO$_2$ reaching a maximum desorption rate at $\sim140$
K during water crystallisation and at 160 K when water starts to
sublimate, allowing trapped volatiles to desorb \citep{Collings2003,
  Collings2004}. In fact, this is observed in Solar System comets,
where the composition of their coma changes from CO- to H$_2$O-driven
near 2.5 AU, where the comet surface temperatures reach $\sim150$~K
\citep[e.g., Comet Hale-Bopp (C/1995 O1) and others,][]{Biver1997,
  Ootsubo2012}.
%% These models did not consider the presence of CO gas trapped nor
%% CO$_2$ ice in comets, which could increase the production rate of
%% gas and dust at larger radii ($\sim20$ AU) due to its lower
%% sublimation temperature}
Therefore, CO could be released in situ explaining the tentative
detection without the need of being released in collisions and
viscously expand from close in regions where the hot dust is. It is
also worth noting that the location at which these temperatures are
reached, depend also on the surface physical properties of the icy
material, e.g. if it is covered by ice or dust, on the rotation rates
and if volatiles are being released from small grains or not, and they
can even vary significantly across the surface of big comets
\citep[e.g., comet
  67P/Churyumov-Gerasimenko,]{Choukroun2015}. Numerical simulations by
\cite{Marboeuf2016} have explored the thermo-physical evolution of
comets and production rates of H$_2$O gas and dust from their surface,
finding for example that the H$_2$O gas and dust production rates
increase exponentially from tens of AU down to 6 AU for $L_\star=5
L_\odot$. Therefore, the CO location and radial distribution, together
with thermo-physical models of comets including other molecule species
such as CO and CO$_2$, can give clues about the nature of these
exocomets and composition.

The latter scenario fits with the hypothesis of icy material being
passed from the outer belt to the inner regions. The non detection of
CO closer in could be explained if the icy material gets depleted of
volatiles before reaching the inner regions where the hot dust
is. This would imply that the timescale at which material loses its
volatile content is shorter than the timescale at which material
migrates from $\sim20$ to $\sim10$ AU, i.e. $\lesssim 2\times10^4$~yr
(using the lower limit on the migration rate for a surface density of
solids proportional to $r^{-1}$ derived in
Sec. \ref{dis:outtoin}). Alternatively, the non-detection could be
also explained by the CO lifetime getting shorter for smaller radii as
the UV stellar radiation starts to dominate in the photodissociation
process, reducing the amount of CO gas. Moreover, we find that the CO
destruction rate, which is $\sim3\times10^{-9}$~M$_\oplus$~yr$^{-1}$,
is of the same order as the hot dust mass loss rate. This implies that
the material from the outer belt passed in is highly rich in CO or
CO$_2$ ($\sim$50\% of mass in CO) or that only a fraction of the
material that reaches 20 AU continues its way in to collide where the
hot dust is.

One potential problem with this scenario is that if the CO present in
the disc is produced in steady state, i.e. constantly released at a
rate of $\sim3\times10^{-9}$~M$_\oplus$~yr$^{-1}$, it would imply that
over a timescale of 1.4 Gyr the outer belt has lost $\sim$4 M$_\oplus$
of CO or 40~M$_\oplus$ in planetesimals assuming a CO mass fraction of
10\% in solids \citep[value consistent with Solar system
  comets,][]{Mumma2011}. This value is of the order of the total
amount of solids initially present at the outer belt location if we
extrapolate the minimum mass solar nebula \citep{Weidenschilling1977}
surface density profile (30~M$_\oplus$), and similar to the initial
Kuiper belt mass in the Nice model
\citep[35~M$_\oplus$,][]{Gomes2005}. However, this assumes that the
amount of CO gas present at 20 AU is in steady state, which is not
necessarily the case. For example, if the hot dust is fed by particles
with a wide size distribution, small grains would be fed continuously,
whereas bigger particles which contain most of the mass would behave
stochastically as the number is much lower. Therefore, we could be
witnessing a rare event in this system. This is similar to the
stochastic accretion proposed to explain the pollution on white dwarfs
\citep{Wyatt2014}. If the tentative CO detection is confirmed and its
distribution resolved, it would help to constrain better how the hot
dust is being fed from material originated in the outer belt.

\section{Conclusions}
\label{sec:conclusions}
% resolved the system and width and CO

In this paper we have presented the first ALMA observations of the
debris disc around \etacorvi \ at 0.88 mm, obtaining the most detailed
image of its outer belt up to date. We detected the outer disc at all
azimuths, with a peak radius of 150~AU and radially spanning over 70
AU or more, consistent with being axisymmetric and with previous
observations.

% study the distribution of mm-dust comparing with different models
In order to obtain estimates of different disc parameters with
uncertainties, we model the emission using a number of four disc
models and compared them with the observed visibilities. The first
model consists of a simple belt with radial and vertical Gaussian mass
distributions. We found the outer belt density distribution peaks at
$150\pm3$ AU, with a radial FWHM of $44\pm6$~AU. This model gives the
best fit to the observations with a total flux of $13\pm1$~mJy. The
second model consists of the expected profile from a self-stirred
disc. We found that the derived disc parameters in the self-stirred
scenario imply an unphysically high range of surface density of solids
in the primordial disc. However, self-stirring could still be the case
if the outer belt was initially narrow or truncated by a planet.

% no LHB
Because of the comet-like composition of the hot dust and its short
lifetime, it has been suggested that it is being fed from the outer
belt. Several mechanisms could be responsible of such
delivery. Previous observations ruled out all except three, one of
which proposes that the system is going through LHB-like instability.
We compared simulated observations using as input one of the LHB
models, scaled to the size of the \etacorvi \ outer belt, finding that
the surface density radial profile from LHB simulations is too broad
compared to the \etacorvi \ outer belt. Moreover, a double power-law
fit to the disc surface density indicates that it must increase
steeply from the inner regions to the outer belt, in contradiction
with a highly scattered belt produced after an instability.  We also
fit an eccentric disc model, finding that the disc is consistent with
being circular and with a $3\sigma$ upper limit for the belt
eccentricity of 0.05. Therefore, we conclude that the outer belt is
probably in a stable configuration on Gyr timescales, where a chain of
planets is scattering in material from the outer belt.

Although the exact mechanism or planetary configuration scattering
material inwards is unknown, we placed upper limits on any millimetre
extended emission from dust inside the cavity based on the measured
visibilities. These limits can be used in the future to place
constraints on the distribution of solids if solid material is indeed
being transported from the outer belt to the inner most regions of the
system.

We searched for any CO gas that could be present in the disc. Although
we did not detect gas in the outer belt or co-located with the hot
dust, we present a tentative detection of CO gas around $\sim20$
AU. Considering non-LTE effects, we derived a CO gas mass estimate
based on a tentative CO detection at $\sim20$~AU and $3\sigma$ upper
limits according to the non-detections in the outer belt and inner
regions. Regarding the origin of the putative CO gas, we find that it
could be released in situ from icy bodies when crossing an ice line,
e.g. CO$_2$ sublimates at $\sim80$ K, which could then release trapped
CO or produce it via photodissociation. Alternatively, CO gas trapped
in porous H$_2$O ice could be released following the crystallisation
of water or its sublimation when reaching 140~K or 150 K,
respectively. This scenario is consistent with the hypothesis that
material being transported from the outer belt to the location of the
hot dust and suggests that we could be observing the system with a
particularly high activity after a recent event. It is unlikely that
the observed CO gas is being produced within a few AU and viscously
expanding outwards up to 20 AU, as timescales for photodissociation
are shorter than viscous evolution.

%% If icy planetesimals can reach the location of the hot dust before
%% loosing its volatile content, CO could be released in a collisional
%% cascade is indeed released in situ. Both scenarios require that a slow
%% migration of solids through the ice line or fast sublimation of ices
%% to lose their ice content before reaching the location of the hot
%% dust.

% place constrains on putative planets in the system
Finally, based on these new observations and the estimates of the
inner edge of the disc and disc eccentricity we put some constraints
on any planet that is responsible for truncating the disc at $\sim$106
AU in a timescale shorter than the age of the system, inducing a
forced eccentricity lower than 0.05, and with a mass and semi-major
axis such that it scatters material from the outer disc that can move
inwards without being ejected from the system on timescales of the
order of the age of the system. Under these restrictions, we find that
such planet should have a semi-major axis around 75-100 AU, a mass
between 3-30 $M_{\oplus}$ and an eccentricity lower than 0.08.

Therefore, we proposed the following global scenario: volatile-rich
solid material formed in the outer belt is being passed in via
scattering with a chain of planets in the system. This icy material
starts to sublimate and loses part of its volatiles when crossing
specific ice line(s), explaining the CO at $\sim$20~AU. The chain of
planets should have a mass distribution close to flat between
3-30~$M_{\oplus}$ to maximise the mass flux into the inner regions
where the hot dust lies. Finally, the inflowing material feeds an in
situ collisional cascade or collides with a planet with a mass of 4-10
M$_\oplus$ located at $\sim3$~AU (sweet spot of the system) releasing
large amounts of debris and resulting in an asymmetric structure,
consistent with LBTI observations and the observed spectroscopic
features. This can be tested by: 1) confirming the CO detection and
resolving both in radius and azimuth the distribution of gas in the
disc with deeper ALMA observations; 2) detecting any extended dust
emission inside the cavity combining ALMA and ACA or with SMA
observations reaching a higher sensitivity on large scales; 3) follow
up LBTI observations to constrain the hot dust distribution which
should remain constant and asymmetric in a giant impact scenario; 4)
RV upper limits of planets that discard the presence of an ejector
within 3 AU ($\gtrsim0.1$~M$_\mathrm{Jup}$); 4) direct imaging to
search for outer planets, which at 10 M$_\oplus$ would be challenging
to detect directly, but could have an enhanced brightness if
surrounded by dust \citep{Kennedy2011}; 5) N-body simulations tailored
to the system.

\section*{Acknowledgements}

We thank Pablo Roman and Simon Casassus for providing us the tool
uvsim to simulate model visibilities. This paper makes use of the
following ALMA data: ADS/JAO.ALMA\#2012.1.00385.S. ALMA is a
partnership of ESO (representing its member states), NSF (USA) and
NINS (Japan), together with NRC (Canada) and NSC and ASIAA (Taiwan)
and KASI (Republic of Korea), in cooperation with the Republic of
Chile. The Joint ALMA Observatory is operated by ESO, AUI/NRAO and
NAOJ. MCW, LM, AB and QK acknowledge the support of the European Union
through ERC grant number 279973. LM also acknowledges support by STFC
through a graduate studentship. The work of O.P. is supported by the
Royal Society Dorothy Hodgkin Fellowship. GMK is supported by the
Royal Society as a Royal Society University Research Fellow.

%%%%%%%%%%%%%%%%%%%%%%%%%%%%%%%%%%%%%%%%%%%%%%%%%%

%%%%%%%%%%%%%%%%%%%% REFERENCES %%%%%%%%%%%%%%%%%%

% The best way to enter references is to use BibTeX:

\bibliographystyle{mnras}
\bibliography{SM_pformation} % if your bibtex file is called example.bib

%%%%%%%%%%%%%%%%%%%%%%%%%%%%%%%%%%%%%%%%%%%%%%%%%%

%%%%%%%%%%%%%%%%% APPENDICES %%%%%%%%%%%%%%%%%%%%%

%% \appendix

%% \section{Some extra material}

%% If you want to present additional material which would interrupt the flow of the main paper,
%% it can be placed in an Appendix which appears after the list of references.

%%%%%%%%%%%%%%%%%%%%%%%%%%%%%%%%%%%%%%%%%%%%%%%%%%

% Don't change these lines
\bsp	% typesetting comment
\label{lastpage}
\end{document}